\documentclass[traditabstract]{aa} 

\usepackage{ graphicx , txfonts , natbib , gensymb , setspace , subfigure , dcolumn , booktabs , xfrac , color, lscape, hyperref }

\begin{document}

\titlerunning{Investigating the variability in a low-mass, pre-main sequence eclipsing binary}

\title{CoRoT\,223992193: Investigating the variability in a low-mass, pre-main sequence eclipsing binary with evidence of a circumbinary disk}

\author{E. Gillen \inst{1,2}
          \and S. Aigrain \inst{2} 
          \and C. Terquem \inst{2}
          \and J. Bouvier \inst{3} 
          \and S. H. P. Alencar \inst{4} 
          \and D. Gandolfi \inst{5,6}
          \and J. Stauffer \inst{7}
          \and A. Cody \inst{8}
          \and L. Venuti \inst{3,9}
          \and P. Viana Almeida \inst{4} 
          \and G. Micela \inst{9}
          \and F. Favata \inst{10} 
          \and H. J. Deeg \inst{11,12}}

   \institute{Astrophysics Group, Cavendish Laboratory, J.J. Thomson Avenue, Cambridge CB3 0HE, UK \\email: \href{mailto:ecg41@cam.ac.uk}{{\tt ecg41@cam.ac.uk}}
        \and Sub-department of Astrophysics, Department of Physics, University of Oxford, Keble Road, Oxford, OX1 3RH, UK 
        \and Univ. Grenoble Alpes, CNRS, IPAG, F-38000 Grenoble, France
        \and Departamento de F\'{i}sica - ICEx - UFMG, Av. Ant\^{o}nio Carlos, 6627, 30270-901, Belo Horizonte, MG, Brazil 
	\and Dipartimento di Fisica, Universit\`a di Torino, via P. Giuria 1, 10125 Torino, Italy
	\and Landessternwarte K\"onigstuhl, Zentrum f\"ur Astronomie der Universit\"at Heidelberg, K\"onigstuhl 12, 69117 Heidelberg, Germany  
        \and Spitzer Science Center, California Institute of Technology, 1200 E California Blvd., Pasadena, CA 91125, USA
        \and NASA Ames Research Center, Moffet Field, CA 94035, USA
       	\and INAF -- Osservatorio Astronomico di Palermo, Piazza del Parlamento 1, 90134, Palermo, Italy
        \and European Space Agency, 8-10 rue Mario Nikis, 75738 Paris Cedex 15, France 
        \and Instituto de Astrof\'\i sica de Canarias, C. Via Lactea S/N, E-38205 La Laguna, Tenerife, Spain
        \and Universidad de La Laguna, Dept. de Astrof\'\i sica, E-38206 La Laguna, Tenerife, Spain
         }

   \date{Received \ldots; accepted \ldots}


\abstract{

CoRoT\,223992193 is the only known low-mass, pre-main sequence eclipsing binary that shows evidence of a circumbinary disk. The system displays complex photometric and spectroscopic variability over a range of timescales and wavelengths. Using two optical CoRoT runs from 2008 and 2011/2012 (spanning 23 and 39 days), along with infrared \emph{Spitzer} 3.6 and 4.5\,$\mu$m observations (spanning 21 and 29 days, and simultaneous with the second CoRoT run), we model the out-of-eclipse light curves, finding that the large scale structure in both CoRoT light curves is consistent with the constructive and destructive interference of starspot signals at two slightly different periods. Using the $v\sin i$ of both stars, we interpret this as the two stars having slightly different rotation periods: the primary is consistent with synchronisation and the secondary rotates slightly supersynchronously. Comparison of the raw 2011/2012 light curve data to the residuals of our spot model in colour-magnitude space indicates additional contributions consistent with a combination of variable dust emission and obscuration. 
There appears to be a tentative correlation between this additional variability and the binary orbital phase, with the system displaying increases in its infrared flux around primary and secondary eclipse.
We also identify short-duration flux dips preceding secondary eclipse in all three CoRoT and \emph{Spitzer} bands. We construct a model of the inner regions of the binary and propose that these dips could be caused by partial occultation of the central binary by the accretion stream onto the primary star.
Analysis of 15 H$\alpha$ profiles obtained with the FLAMES instrument on the Very Large Telescope reveal an emission profile associated with each star. The majority of this is consistent with chromospheric emission but additional higher velocity emission is also seen, which could be due to prominences. However, half of the secondary star's emission profiles display full widths at 10\% intensity that could also be interpreted as having an accretion-related origin. In addition, simultaneous $u$ and $r$-band observations obtained with the MEGACam instrument on the Canada France Hawaii Telescope reveal a short-lived u-band excess consistent with either an accretion hot spot or stellar flare. 
The photometric and spectroscopic variations are very complex but are consistent with the picture of two active stars possibly undergoing non-steady, low-level accretion; the system's very high inclination provides a new view of such variability.

}

  
   \maketitle



\section{Introduction}

Classical T\,Tauri stars (CTTS) are young solar-type stars ($M < 2M_{\odot}$), which accrete material from their circumstellar disks. They display both photometric and spectroscopic variability over a range of timescales (from hours to years) and wavelengths (from ultraviolet to infrared). This variability is inferred to result from processes at and near the stellar surface and can be categorised into intrinsically stellar and accretion-related processes \citep[e.g.][]{Bouvier07,Cody14}. Weak-lined T\,Tauri stars (WTTS) represent young systems that are not actively accreting and display only stellar variability \citep[e.g.][]{Grankin08}.

Intrinsic stellar variations are caused by surface inhomogeneities arising from cool, magnetically active starspots, which cause photometric and spectroscopic modulation due to the star's rotation. Accretion-related variations arise from the infall of material from the circumstellar disk onto the star which, in the most commonly accepted paradigm, is mediated by the stellar magnetic field. Zeeman measurements indicate typical surface field strengths of order a few kilogauss \citep[e.g.][]{Symington05,Donati08}, which is strong enough to disrupt the inner disk flow at a few stellar radii, truncating the disk and funnelling material towards the star along the magnetospheric field lines. The structure and evolution of these accretion columns are governed by the field configuration, the inclination between the stellar rotation and magnetic axes, and the mass accretion rate \citep{Romanova03,Romanova08,Long07,Long08}. As material approaches the stellar surface it reaches near free-fall velocities, dissipating its kinetic energy in a shock at the stellar surface and heating the immediate area.

In this context, an actively-accreting TTS could display photometric variability arising from cool and hot spot modulation, changing mass accretion rates and variable extinction \citep[e.g.][]{Venuti15}. Extinction can result from obscuration of the central star by the inner disk wall, either due to a warp, as in AA\,Tau-like objects \citep[e.g.][]{Bouvier99,Fonseca14}, or possibly due to changing disk height \citep[e.g.][]{Flaherty10,Espaillat11}. Spectroscopically, such a star would display variable permitted emission lines whose profiles might show evidence of hot spots, accretion columns and disk winds, and in a less well-ordered accretion framework, emission from material in the stellar magnetosphere \citep{Alencar12}. High mass accretion rates can also drive stellar winds and jets, which can be investigated through forbidden emission line profiles \citep{Shang02}. 

TTS variability is ubiquitous and has been studied since their discovery by \citet{Joy45}. In recent decades, most ground-based studies have focussed on optical and near-infrared (near-IR) photometry \citep[e.g.][]{Bouvier93,Herbst94,Makidon04,Grankin07,Parks14}. Recent observations with the \emph{Spitzer Space Telescope} and \emph{Herschel} detected flux variations in disk-bearing stars in the mid and far-IR \citep[e.g.][]{Morales-Calderon11,Espaillat11,Billot12}. Combining the available data with advances in modelling of the star-disk interaction and inner disk dynamics has furthered our understanding of the underlying physics but has yet to unambiguously distinguish between plausible physical scenarios, in part due to the sparseness and non-simultaneity of the data \citep{Flaherty10,Romanova11,Romanova13}.

Further progress may come from the ongoing YSOVAR project \citep[Young Stellar Object Variability;][]{Morales-Calderon11,Rebull14}, which monitors a dozen young clusters at high cadence with \emph{Spitzer}/IRAC at 3.6 and 4.5 $\mu$m. In addition, the recent Coordinated Synoptic Investigation of NGC 2264 (CSI 2264) comprises the most extensive continuous simultaneous multi-band photometric and spectroscopic dataset ever compiled for a young star forming region. It is introduced in detail in \citet{Cody14} who analyse simultaneous optical CoRoT and IR \emph{Spitzer} photometry of 162 disk bearing stars to classify their variability into seven morphological variability classes, which they argue represent different physical mechanisms and geometric effects. Surprisingly, they also find that optical and IR variability is not correlated in the majority of cases. Additional insights have come from analysing AA\,Tau analogs and their cousins with deep, aperiodic flux dips \citep{McGinnis15}, as well as a similar, but distinct, class of CTTS that displays narrow, periodic flux dips \citep{Stauffer15}. Furthermore, a new morphological class of CTTS was reported by \citet{Stauffer14} whose light curves are dominated by short-duration accretion bursts.

The vast majority of variability studies to date have focussed on single T\,Tauri stars in clusters. As a significant fraction of stars form in binaries or higher order systems \citep{Duchene13} it is important to understand both the physical processes at play and their effect on the system's formation and early evolution. In well-separated, accreting binary systems one might expect individual circumstellar disks around each star, as well as a circumbinary disk around both, separated by a region of very low density, which arises from the transfer of angular momentum from the binary to the circumbinary disk \citep{Artymowicz94}. Material streams through this central cavity from the circumbinary disk onto the circumstellar disks, which in turn accrete onto the stars \citep{Artymowicz96,Gunther02}. One might therefore expect any of the aforementioned variability to be present for each star as well as additional contributions from the accretion streams and the inner regions of the circumbinary disk. In close-separation binaries substantial circumstellar disks may be prohibited due to tidal truncation by the other star \citep{Paczynski77,Papaloizou77} but accretion can still take place via the accretion streams. Due to the greater geometric complexity of binary systems, characterising the physical origins of their variability is more challenging than for single stars, and hence one needs well-characterised, benchmark systems to use as test-beds.

One such system is CoRoT\,223992193, which is a low-mass, pre-main sequence eclipsing binary (PMS EB) member of the $\sim$3 Myr old NGC\,2264 star-forming region. It was discovered during a 23 day observation of NGC 2264 by the CoRoT space mission in 2008. An initial analysis, presenting the fundamental parameters of the two stars and modelling of the system's spectral energy distribution (SED) is given in \citet[][hereafter Paper 1]{Gillen14}.

To briefly recap, the EB comprises two M dwarfs, which have component masses of $M_{\rm{pri}} = 0.67 \pm 0.01$ and $M_{\rm{sec}} = 0.495 \pm 0.007$ $M_{\odot}$, and corresponding radii of $R_{\rm{pri}} = 1.30 \pm 0.04$ and $R_{\rm{sec}} = 1.11^{+0.04}_{-0.05}$ $R_{\odot}$. Their orbit is circular with a period of $3.8745745 \pm 0.0000014$ days and constant separation $a = 10.92 \pm 0.06$ $R_{\odot}$. Comparison with a selection of PMS stellar evolution models indicate an apparent age of $\sim$4-5 Myr, consistent with the age of the cluster \citep[Paper 1;][]{Stassun14}.

\begin{figure*}
\centering
\includegraphics[width=0.49\linewidth]{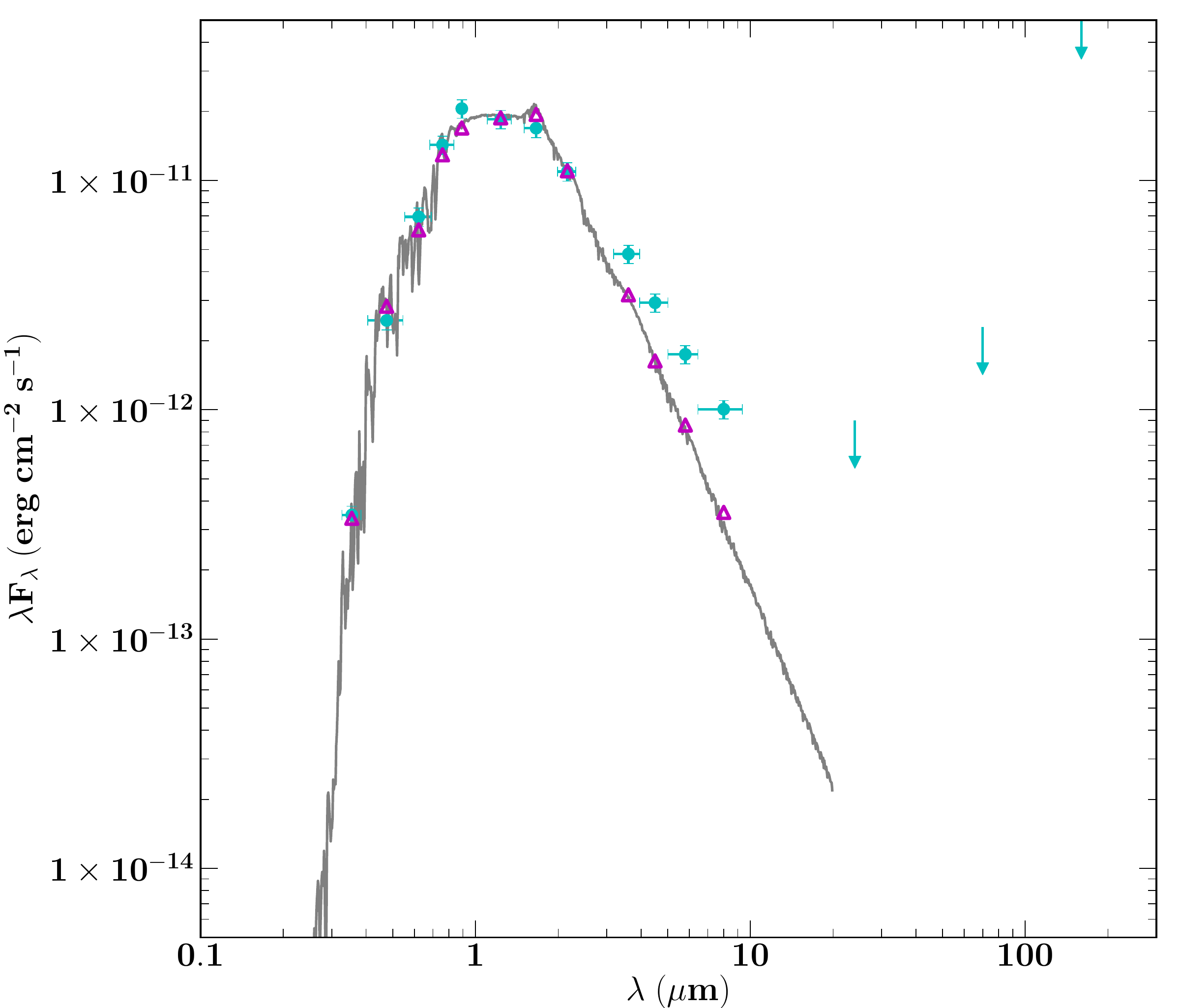} \hfill
\includegraphics[width=0.49\linewidth]{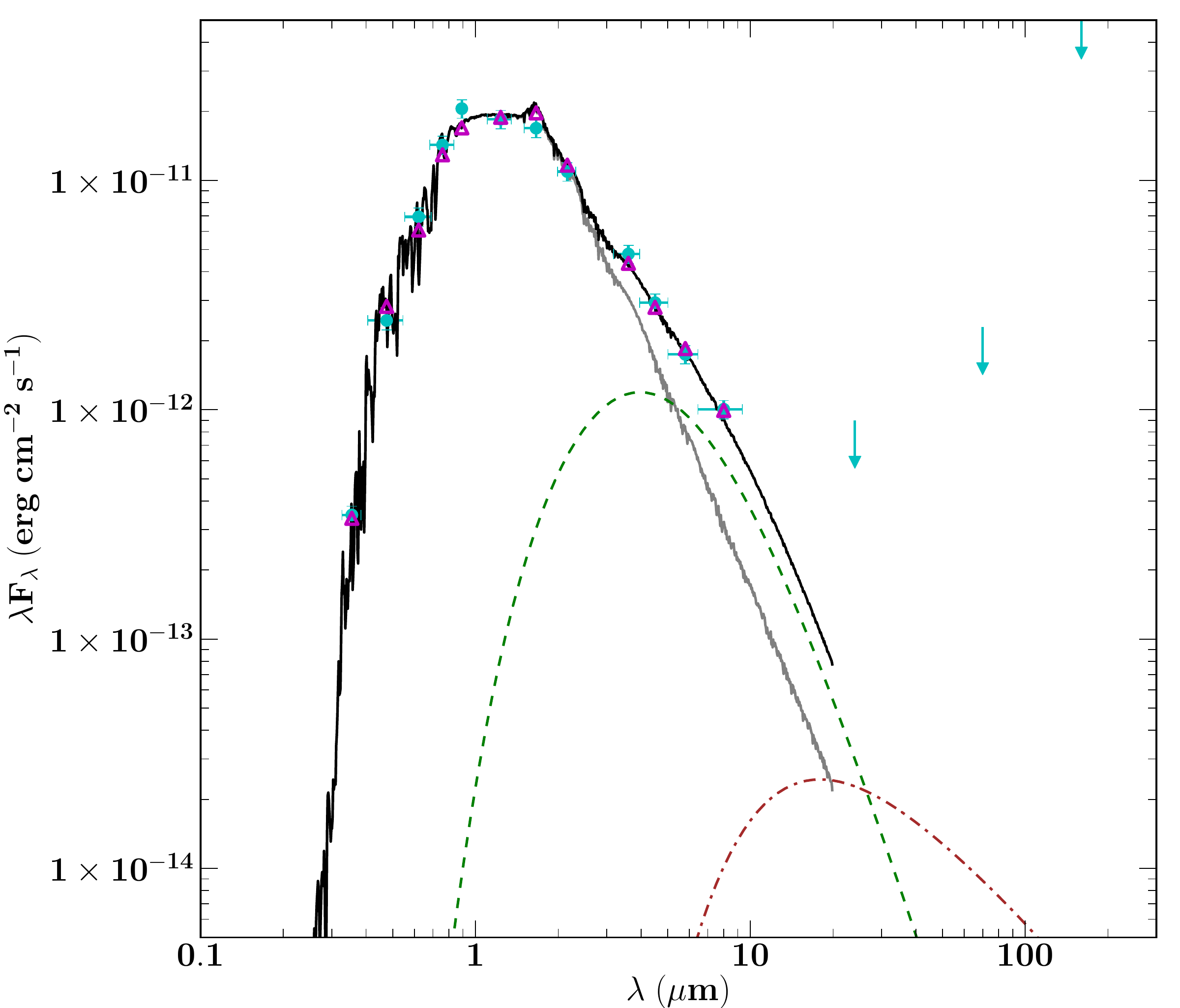}
\caption{Spectral energy distribution of CoRoT\,223992193 (cyan points) with upper limits in the far-infrared (cyan arrows). This figure is identical to Fig.~13 of \citet[][Paper 1]{Gillen14}; it is reproduced here for clarity. Left: the grey line and magenta triangles show the best-fit model using two naked stellar photospheres. Right: the black line and magenta triangles show the best-fit two naked photospheres model with a small amount of hot dust in the inner cavity of the circumbinary disk. The stellar and hot dust terms are shown separately as the grey solid and dashed green lines, respectively. Also shown for completeness, but not used in the fit, is the expected emission from a razor-thin circumbinary disk extending down to 22\,$R_{\odot}$ (brown dot-dashed line), which is illuminated by the central star and heated by the gravitational potential energy released from accretion with $\dot{M} = 10^{-11} M_{\odot}$\,yr$^{-1}$.}
\label{sed}
\end{figure*}

\begin{figure}
   \centering
   \includegraphics[width=\linewidth]{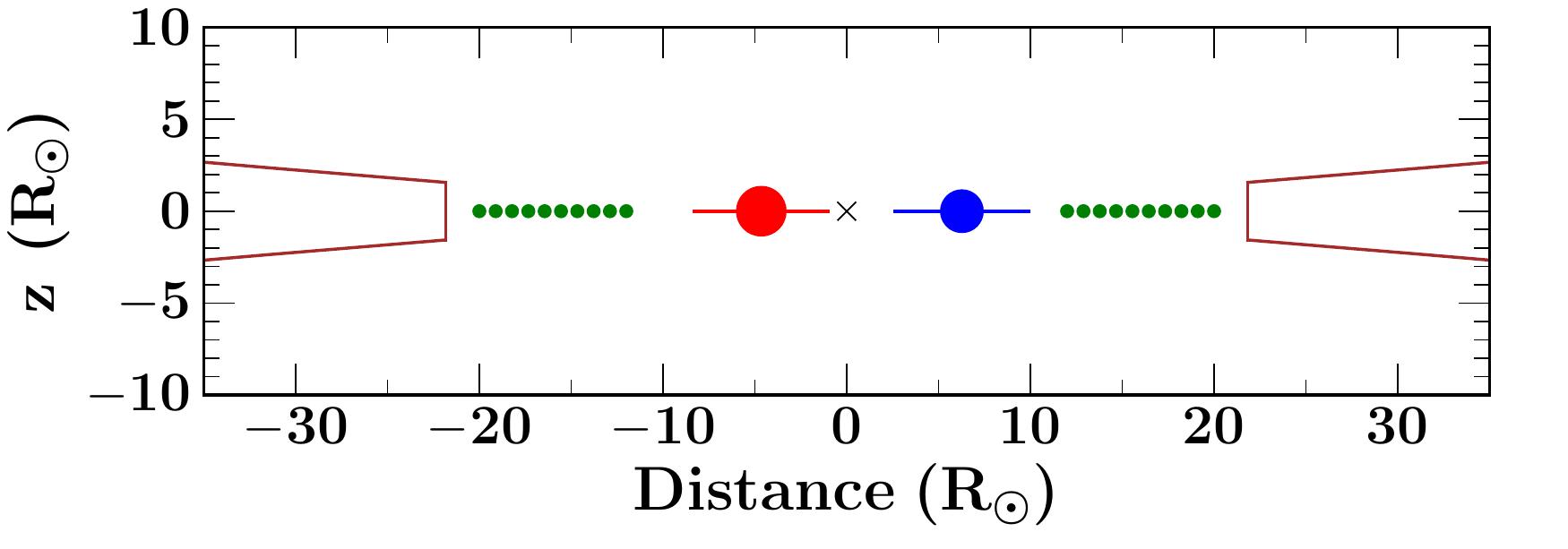}
   \caption{Schematic representation of the proposed system geometry, showing distance from the centre of mass against height above the system plane (z). This figure is identical to Fig.~14 of \citet[][Paper 1]{Gillen14}; it is reproduced here for clarity. The primary and secondary stars, along with their circumstellar disks (truncated to a third of the binary separation), are shown in red and blue respectively (the circumstellar disks are shown for completeness but there is no direct evidence for their presence). The sizes and separations of the two stars are to scale. The circumbinary disk (brown) has its inner radius truncated at twice the binary separation. The green dots indicate the general location of dust lying within the inner cavity of the circumbinary disk, such as one could expect to find in accretions streams.}
    \label{sys_view}
\end{figure}

This system is particularly interesting because it shows evidence for a circumbinary disk due to a mid-IR excess in its SED, which is shown in Fig.~\ref{sed} (reproduced from Fig.~13 of Paper 1 for clarity). \citet{Sung09} classify the system as Class II/III based on the \emph{Spitzer}/IRAC magnitudes. The reader is directed to section 4.5 of Paper 1 for a full discussion; here, we briefly recap the salient points. We initially modelled the SED as the sum of two naked stellar photospheres (Fig.~1, left panel) but could not reproduce the slope of the SED in the \emph{Spitzer}/IRAC bands. Given the system's youth, we tested whether this excess could be due to extended dust emission in the vicinity of the two stars. We found that the dynamics and geometry of the system preclude large, stable circumstellar disks, and furthermore, we require dust at cooler temperatures than would be expected in these disks to explain the SED. We found that the mid-IR excess is consistent with thermal emission from a small amount of dust located in the central cavity of a circumbinary disk. In our simple model, this dust extends from $\sim$5 -- 32 $R_{\odot}$ with corresponding temperatures of $\sim$1450 to 600 K (Fig.~\ref{sed}, right panel). For dust to still be present in the cavity at $\sim$4 Myr requires replenishment and we thereby inferred the presence of a circumbinary disk as the most natural means of doing so. Unfortunately, the lack of high resolution far-IR data meant we could only place relatively weak upper limits on the circumbinary disk itself (limited by the background nebulosity). A schematic representation of the proposed system geometry is shown in Fig.~\ref{sys_view} (reproduced from Fig.~14 of Paper 1). 

Further circumstantial evidence for the circumbinary disk hypothesis comes from the fact that the 2008 CoRoT light curve displays large amplitude, rapidly evolving out-of-eclipse (OOE) variations that are difficult to explain with simple star spot models alone. Given the high inclination of the system ($i = 85.09^{+0.16}_{-0.11} \,\degree$) and the presence of dust in the cavity of the circumbinary disk, we suggested the possibility that some of the OOE variations seen in the light curve could be due to occultation of one or both stars by material located at the inner edge, or in the central cavity, of the circumbinary disk.

CoRoT\,223992193 is an ideal candidate for studying variability in a young, close-separation binary for two reasons: 
a) the stellar properties and system geometry are known to a precision unattainable for non-eclipsing systems, and 
b) we have obtained continuous simultaneous multi-band photometric and spectroscopic observations (as part of CSI\,2264), which allow us to probe the variability over a range of timescales and wavelengths.

In this paper we address the following question: what are the physical origins of both the photometric and spectroscopic variability seen in CoRoT\,223992193. 
In section \ref{observations} we give details of the photometric and spectroscopic observations. In section \ref{phot_var_sec} we model the multi-band light curves and discuss plausible physical origins for the different types of variability. In section \ref{spec_var_sec} we model the system's H$\alpha$ profiles and search for other emission lines, and in section \ref{var_consistency} we perform consistency checks between the photometric and spectroscopic data. Finally, we present a model of the inner regions of the binary and propose a possible scenario to explain the observed short-duration flux dips in section \ref{sdfd}, before concluding in section \ref{conclusions}.


\section{Observations}
\label{observations}

The NGC\,2264 star forming region was observed by the CoRoT space mission for 23 days in March 2008. Almost four years later, it was re-observed by 15 ground and space-based telescopes during December 2011 -- March 2012 comprising the CSI\,2264 campaign. Here we focus on those observations relevant to the characterisation of CoRoT\,223992193.

In Paper 1 we characterised the stellar orbits and determined the fundamental parameters using the 2008 CoRoT observations, as well as spectroscopy obtained with the FLAMES multi-object spectrograph on the Very Large Telescope (VLT) at Paranal, Chile (obtained as part of CSI\,2264) and the Intermediate dispersion Spectrograph and Imaging System (ISIS) on the \emph{William Herschel} Telescope situated on La Palma. 

In this paper we focus on the 2008 CoRoT observations and selected CSI\,2264 data to investigate the variability in CoRoT\,223992193. The CSI\,2264 photometric observations primarily consist of simultaneous CoRoT and \emph{Spitzer} light curves. We also use $u$ and $r$-band observations taken with the Megacam instrument on the Canada France Hawaii Telescope (CFHT) situated on Mauna Kea, Hawaii, which were obtained the month after the 2011/2012 CoRoT/\emph{Spitzer} dataset. The spectroscopic observations consist of optical VLT/FLAMES spectra (the same as in Paper 1) and follow-up optical spectra taken with the Fibre-fed Echelle Spectrograph (FIES; \citealt{Frandsen99,Telting14}) on the Nordic Optical Telescope (NOT), situated on La Palma.

\subsection{Photometry}

CoRoT: 
The 2008 CoRoT observations spanned  23.4 days (7 -- 31 March 2008; PI Favata) and comprised the first CoRoT short run in the galactic anticentre direction (SRa01).
The CSI\,2264 CoRoT observations spanned 38.7 days (1 December 2011 -- 9 January 2012; PI Micela) and comprised the fifth CoRoT short run in the galactic anticentre direction (SRa05). CoRoT observations are conducted in a broad 370\,--\,1000\,nm bandpass with a standard cadence of 512\,s, giving 3936 and 6528 photometric data points for the 2008 and 2011/2012 runs, respectively\footnote{Roughly half way through the 2011/2012 run the onboard software automatically changed to high cadence mode (32s); for this work, these data were binned to the standard cadence.}. 

\emph{Spitzer}: The CSI\,2264 \emph{Spitzer}/IRAC observations (PI Stauffer) of CoRoT\,223992193 were conducted in mapping mode with the 3.6 and 4.5\,$\mu$m Warm Mission bands. CoRoT\,223992193 fell within the central IRAC mapping region resulting in near-simultaneous photometry in both bands. The observations spanned 23.5 days (8 December 2011 -- 1 January 2012) for 3.6 $\mu$m and 28.7 days (3 December 2011 -- 1 January 2012) for 4.5 $\mu$m, each with a cadence of $\sim$2 hours. Towards the beginning of the campaign \emph{Spitzer} operated in staring mode for four blocks of $\sim$20 hours each. CoRoT\,223992193 did not fall within the observed region resulting in corresponding gaps early on in the \emph{Spitzer} light curves. Throughout the paper, \emph{Spitzer} fluxes are referred to as 3.6 and 4.5\,$\mu$m and magnitudes as [3.6] and [4.5].

CFHT: The CSI\,2264 CFHT/MegaCam observations (PI Bouvier) consisted of deep $ugri$ mapping as well as $u$ and $r$-band monitoring of the entire NGC\,2264 region over a 14 night period (14 -- 28 February 2012). On each monitoring night, the region was repeatedly imaged with a temporal cadence ranging from 20 mins to 1.5 hours. Each typical observing block consisted of 5 $r$-band exposures followed by 5 $u$-band exposures, each utilising a 5-step dithering pattern, with individual exposure times of 3 and 60s, respectively. Observations conducted in non-photometric conditions were discarded giving 38 $u$-band and 43 $r$-band epochs spread over 11 nights within the two week period.

The reader is directed to Paper 1 for further details on the 2008 CoRoT observations, to \citet{Cody14} for detailed overviews of the observing strategy, data reduction and light curve production for the 2011/2012 CoRoT and \emph{Spitzer} observations, and similarly to \citet{Venuti14} for the CFHT observations.

USNO: Near-IR Cousins I band monitoring of NGC\,2264 was obtained with the USNO 40-inch telescope between 23 November 2011 -- 8 March 2012 (PI Vrba).  Data were not obtained on all nights, but typically 5--20 images were acquired on each observation night. Four fields were observed with the 2048$\times$2048 ``new2K'' CCD, each 23$\times$23 arcmin, with the one including CoRoT\,223992193 centred at RA = 6:41:31.7 and Dec = +09:31:53. Exposure times were typically 300 seconds (although they ranged between 30--900 seconds) and seeing ranged from 1.5 to 8 arcsec. The data were reduced using standard bias subtraction and flatfielding techniques (using dome flatfield images). Differential photometry was then produced relative to seven bright, photometrically non-variable, cluster non-members\footnote{These targets displayed light curves constant to 1\% or better and lacked any evidence for cluster membership.}. This process yielded 859 differential photometric data points for CoRoT\,223992193.

\subsection{Spectroscopy}

VLT\,/\,FLAMES: The CSI\,2264 VLT/FLAMES observations (GO program 088.C-0239(A); PI Alencar) consisted of 15 medium resolution optical spectra obtained over a $\sim$3 month period (4 December 2011 -- 24 February 2012), with both sparsely and densely sampled time intervals. These spectra cover the wavelength range $\sim$6440--6820\AA\ with a resolving power R $\sim$ 17\,000. The reader is directed to Paper 1 for more details of the reduction process (section 2.2) and an example full spectrum (Fig. 5). In this work we focus exclusively on the broad, resolved and highly variable H$\alpha$ emission line.

NOT/FIES: We obtained one spectrum per night for three consecutive nights (4--6 January 2013) on the 2.56-m NOT (PIs Gandolfi and Deeg). The spectra cover the wavelength range 3630--7170\,\AA\ with a resolution R\,$\sim$\,25\,000. We followed the same observing strategy as described in \citet{Gandolfi13} and reduced the data using standard IRAF and IDL routines. Due to the relative faintness of CoRoT\,223992193 (R=15.8), 1-hour integration-time -- under clear and good sky conditions -- yielded a S/N of 5-15 from H$\beta$ to H$\alpha$. This is insufficient to properly investigate the full range of emission lines in this object and perform any meaningful spectral analysis. We used the FIES spectra to search for and classify additional optical emission lines (section \ref{other_em_lines}), and to estimate the projected rotational velocity (section \ref{vsini}).

\begin{figure}[t!]
  \centering  
  \includegraphics[width=\linewidth]{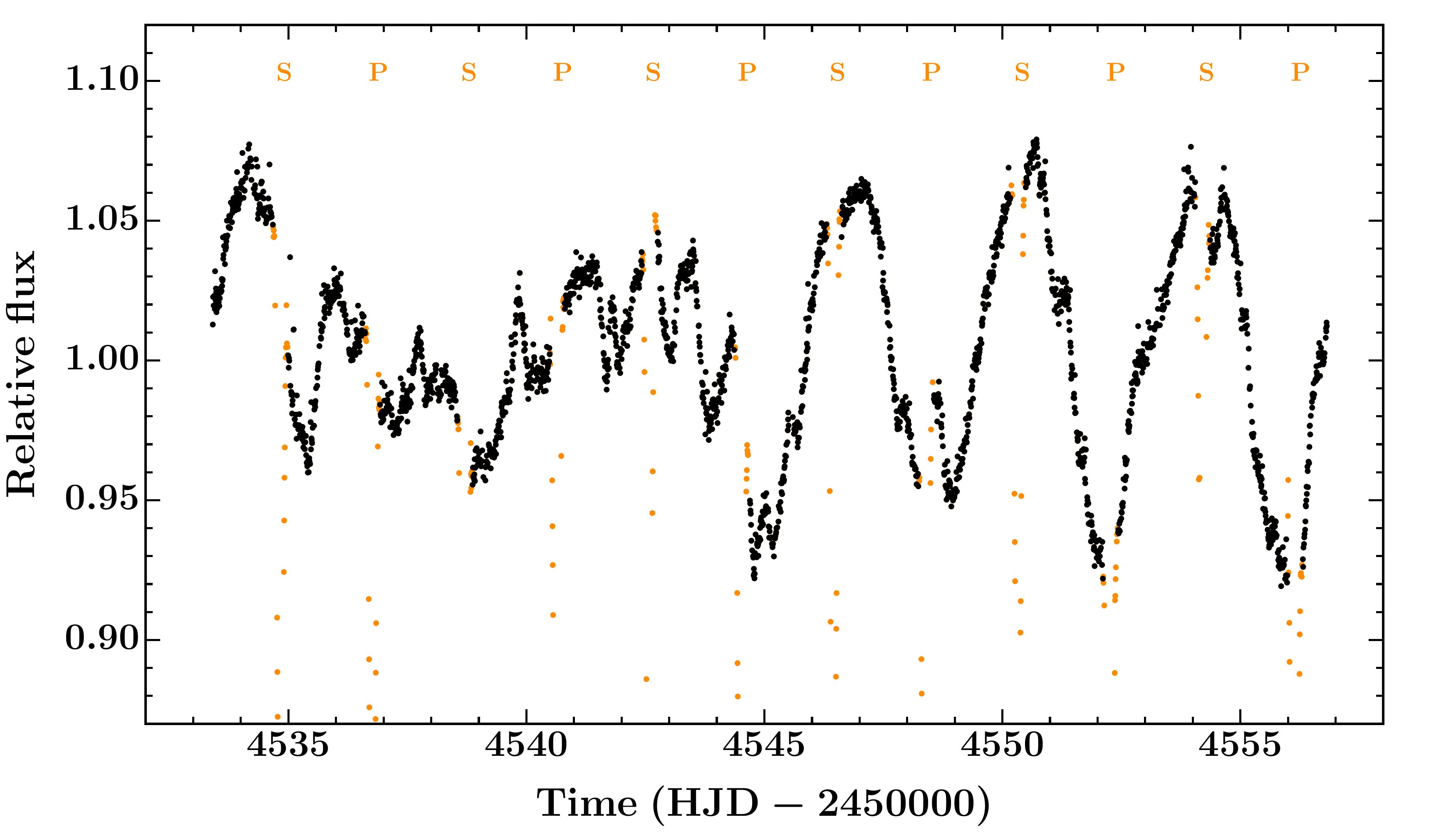}
   \caption{CoRoT light curve from the 2008 observations. The out-of-eclipse data is shown in black and the stellar eclipses in orange (P and S indicate primary and secondary eclipses, respectively).}
   \label{2008_OOE_LC}
\end{figure}

\begin{figure*}[t!]
  \centering  
  \includegraphics[width=\linewidth]{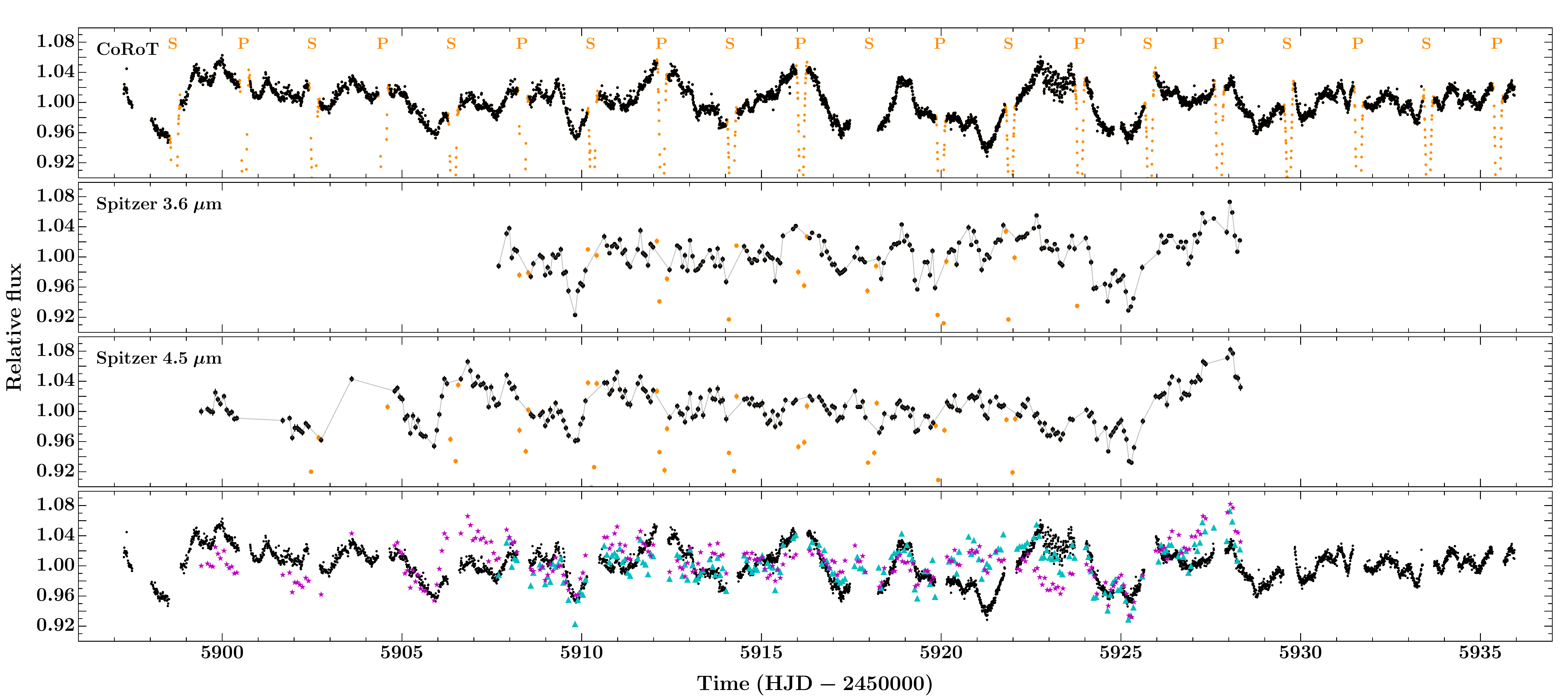}
   \caption{CoRoT and \emph{Spitzer} light curves from the CSI\,2264 campaign. \emph{Top three plots}: optical CoRoT, IR \emph{Spitzer} 3.6\,$\mu$m and \emph{Spitzer} 4.5\,$\mu$m light curves. The out-of-eclipse data is shown in black and the stellar eclipses in orange (P and S indicate primary and secondary eclipses, respectively). To help guide the eye, grey lines join the out-of-eclipse data in the \emph{Spitzer} bands.
   \emph{Bottom}: CoRoT, 3.6 and 4.5\,$\mu$m out-of-eclipse light curves (without errors) over-plotted (black points, cyan triangles and magenta stars, respectively). All plots share common axes.}
   \label{OOE_LCs}
\end{figure*}


\section{Photometric variability}
\label{phot_var_sec}

\subsection{Overall light curve morphology}

\subsubsection{2008 CoRoT light curve}

The 2008 CoRoT light curve is shown in Fig.~\ref{2008_OOE_LC} with the out-of-eclipse (OOE) data in black and stellar eclipses in orange. There are two apparent `regimes' in the OOE variability: one where small amplitude, short-timescale variations (SAVs) dominate the structure (cf. rHJD\footnote{Reduced heliocentric julian date = HJD - 2450000} $\sim$4536--4543), and one where large amplitude, quasi-periodic variations (LAVs) dominate (cf. rHJD $\sim$4545--4557), but where the SAVs are still present.

\subsubsection{Simultaneous 2011/2012 CoRoT and \emph{Spitzer} light curves}				        

The simultaneous optical and IR light curves obtained by CoRoT and \emph{Spitzer} during the 2011/2012 campaign confirm the long-lived nature of the intriguing out-of-eclipse (OOE) variability seen in 2008 (Fig.~\ref{2008_OOE_LC}) and can help us investigate its physical origin(s). The top three plots of Fig.~\protect\ref{OOE_LCs} show the CoRoT and \emph{Spitzer} 3.6 and 4.5 $\mu$m light curves with the OOE data in black and stellar eclipses in orange. The CoRoT light curve displays a similar morphology to that seen in 2008: small amplitude, short-timescale variations (SAVs) dominate the structure between rHJD $\sim$5900--5910 and between 5930--5936, and large amplitude, quasi-periodic variations (LAVs) dominate between rHJD $\sim$5910-5925, but again the SAVs are still present.

The light curve morphologies of the two IR \emph{Spitzer} bands appear to be dominated by the SAVs; the LAVs seen in the CoRoT light curves have a reduced amplitude indicating a relatively warm origin. The most striking common feature in all three bands of the 2011/2012 dataset are short, sharp flux dips, e.g. at rHJD $\sim$ 5910 and 5925, which occur just before secondary eclipse (i.e. at multiples of the binary orbital period) and display different colour signatures throughout the light curves.
There are also differences in behaviour, e.g. at rHJD $\sim$ 5926 the CoRoT flux falls whereas both \emph{Spitzer} fluxes rise. 
The complexity of the observed variations suggests multiple origins.

\subsubsection{CFHT light curves}

\begin{figure}[t!]
  \centering  
  \includegraphics[width=\linewidth]{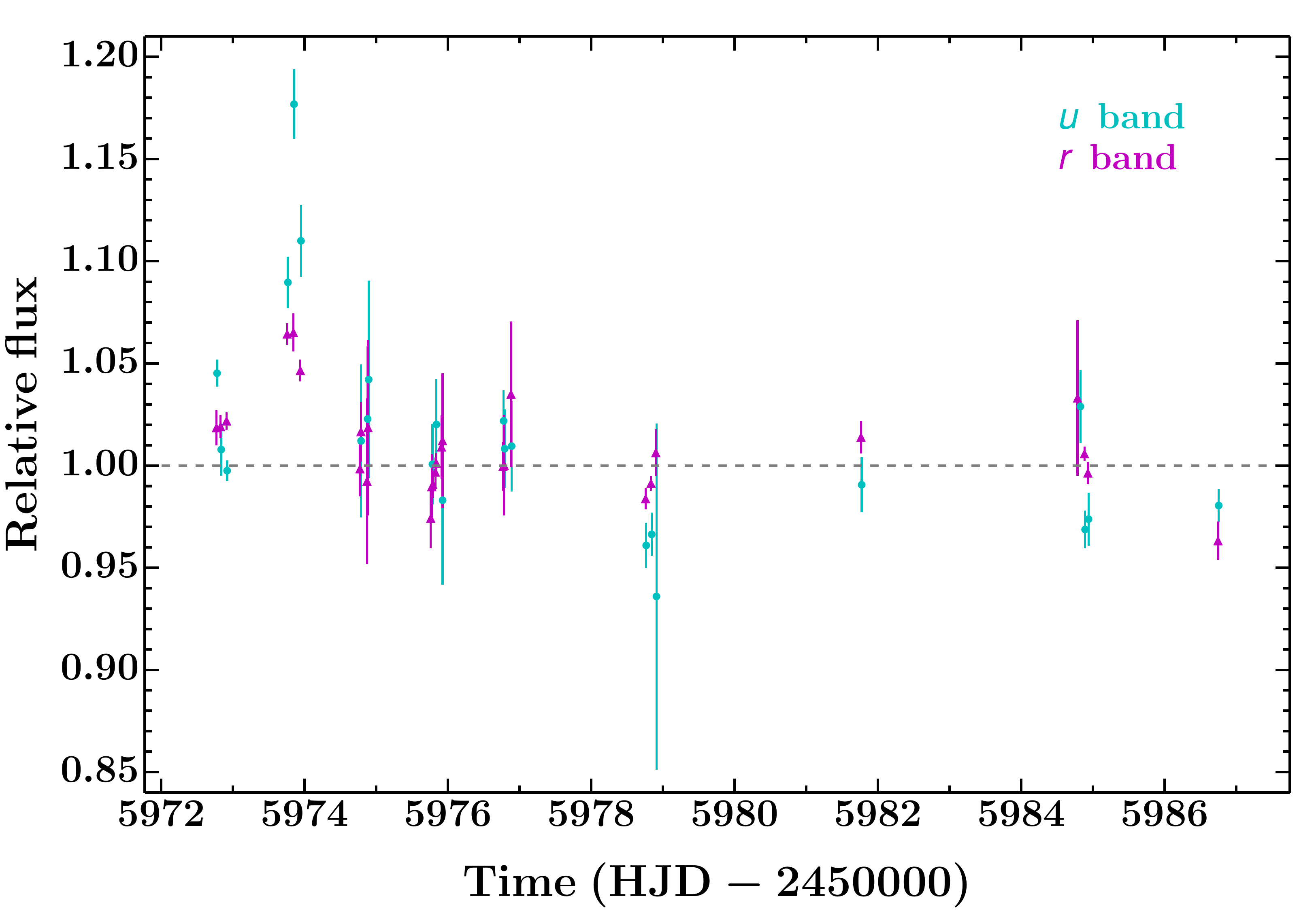}
   \caption{CFHT $u$ and $r$-band out-of-eclipse light curves (cyan circles and magenta triangles, respectively). A short-lived $u$-band excess at rHJD $\sim$ 5974 can clearly be seen, which could be due to either an accretion hot spot or a stellar flare.}
   \label{CFHT_LCs}
\end{figure}

Simultaneous $u$ and $r$-band light curves obtained with CFHT/Megacam the month after the 2011/2012 CoRoT/\emph{Spitzer} run (Fig.~\protect\ref{CFHT_LCs}) show a short-lived $u$-band excess at rHJD $\sim$ 5974. Perhaps the simplest interpretation of this is a stellar flare \citep[e.g.][]{Fernandez04}. However, given that we require dust in the central cavity of the circumbinary disk to explain the SED it is plausible that some of this material accretes onto one or both stars. The $u$-band excess could therefore be explained as a short-lived accretion hot spot resulting from accreting material heating the stellar surface upon impact. Due to the sparsely sampled CFHT data it is difficult to differentiate between these two options from the shape of the $u$-band excess or to place meaningful constraints on the frequency of such events, as we are insensitive to many periods. Nonetheless, this opens up the possibility that some of the variations in the CoRoT and \emph{Spitzer} light curves could be due to short-lived, and potentially recurring, hot spot emission or to stellar flares. We recall that the system's average colours do not show a significant $u$-band excess (Paper 1, section 4.5), implying a low-to-negligible average accretion rate consistent with the majority of the CFHT light curve.

It is important to note that the raw CFHT light curves consist of short runs of 5 consecutive observations and these have been binned using a weighted average to give each data point seen in the figure. Hence, the three points indicating a $u$-band excess are derived from 15 independent observations conducted in photometric conditions.

\subsection{Spot model}
\label{spot_model}

Young, low-mass stars tend to be heavily spotted \citep[e.g.][]{Donati00} and can display photometric modulations not unlike the LAV-dominated parts of the CoRoT light curves. However, changes in such modulations from starspot evolution tend to occur smoothly on timescales of many days to weeks \citep[e.g.][]{Roettenbacher13}, rather than the sudden morphological changes we see at rHJD $\sim$ 4544 in the 2008 CoRoT light curve (when the LAVs appear) and at rHJDs $\sim$ 5910 and 5930 in the 2011/2012 CoRoT light curve (when the LAVs appear and disappear, respectively).

Given that there are two stars in this system, both of which are likely spotted, an appealing solution could be constructive and destructive interference of starspot signals where different active regions have slightly different periods. This could arise due to either different stellar rotation periods (the system is young and not necessarily synchronised) or, even if the stars are synchronised, different latitudes (differential rotation)\footnote{For the latter case, starspots on the same star would be indistinguishable from spots on two stars.}. We therefore sought to understand how much of the large scale structure in the CoRoT and \emph{Spitzer} OOE light curves could be attributed to the constructive and destructive interference of starspot signals.

We set up a simple two-spot model following the formalism of \citet{Dorren87}. Due to the well known degeneracy of spot models we allow only two starspots and opt for a single spot on each star rather than two spots on one star\footnote{Note that these spots should more realistically be seen as a group of many small spots covering a large fraction of the stellar surface around the specified location.}. Before modelling the data we need to: a) determine the flux contributions of each star in each band, b) determine the flux ratios between the spotted and unspotted stellar photospheres and c) mask variations that are clearly not due to spot modulation: 

a) the observed flux in each band is the sum of the stellar and dust components. Our SED modelling indicated that the dust contributes negligible flux in the CoRoT band but makes up a significant fraction at 3.6 and 4.5\,$\mu$m. The stellar flux fractions are 0.62, 0.41 and 0.34 for the primary star in the CoRoT, \emph{Spitzer} 3.6 and 4.5 $\mu$m bands respectively, and 0.38, 0.25 and 0.21 in turn for the secondary.

b) we determine flux ratios using PHOENIX model spectra \citep{Husser13} with a surface gravity of $\log g = 4.0$ and assuming unspotted stellar temperatures of $T_{\rm{pri}} = 3700$\,K and $T_{\rm{sec}} = 3600$\,K (as determined in Paper 1). Limb darkening (LD) coefficients for each band were estimated from \citet{Claret12} assuming the above parameters and solar metallicity \citep[the closest available to the cluster metallicity;][]{King00}.

c) visual examination of the light curves revealed variations which clearly could not be explained by starspot modulation, based on the sharpness and amplitude of their features and/or their colour signatures; these data were masked when fitting the light curves. It was difficult to unambiguously identify such variations in the 2008 CoRoT light curve but was much easier in the 2011/2012 light curves due to the three colour bands. We note that the regions of the light curves satisfying these criteria in the 2011/2012 dataset occur just before secondary eclipse (i.e. separated by multiples of the EB orbital period), which could give clues as to their origin (we discuss this further in section \ref{sdfd}).

We then modelled the 2008 CoRoT light curve and simultaneously modelled the 2011/2012 CoRoT and \emph{Spitzer} 3.6 and 4.5 $\mu$m light curves using the Affine Invariant Markov chain Monte Carlo (MCMC) method as implemented in {\tt emcee} \citep{Foreman-Mackey13}, stepping through the parameter space 50\,000 times with each of 192 `walkers' in both cases. The first 25\,000 steps were discarded as `burn in' and parameter distributions derived from the remainder. The parameters of the fit were the size ($\alpha$), temperature ($T_{\rm{spot}}$), latitude ($\delta$), longitude ($\phi_{0}$) and rotational period ($P_{\rm{rot}}$) of each spot, as well as individual zero points ($F_{\rm{max}}$) and jitter terms ($\sigma$) for each light curve to account for different relative offsets and the fact that the variations are more complicated than a simple two spot model, respectively. In test runs, the individual `walkers' of the MCMC were initialised uniformly from a reasonable section of the parameter space utilising a latin hypercube approach to ensure unbiased start points \citep[e.g.][]{McKay79}; this is a useful approach is cases where the likelihood space is complex. The final runs were initialised from a smaller section of the parameter space based on the posterior distributions of parameters from the test runs, again using a latin hypercube approach.

\begin{figure}[t!]
  \centering
      \includegraphics[width=\linewidth]{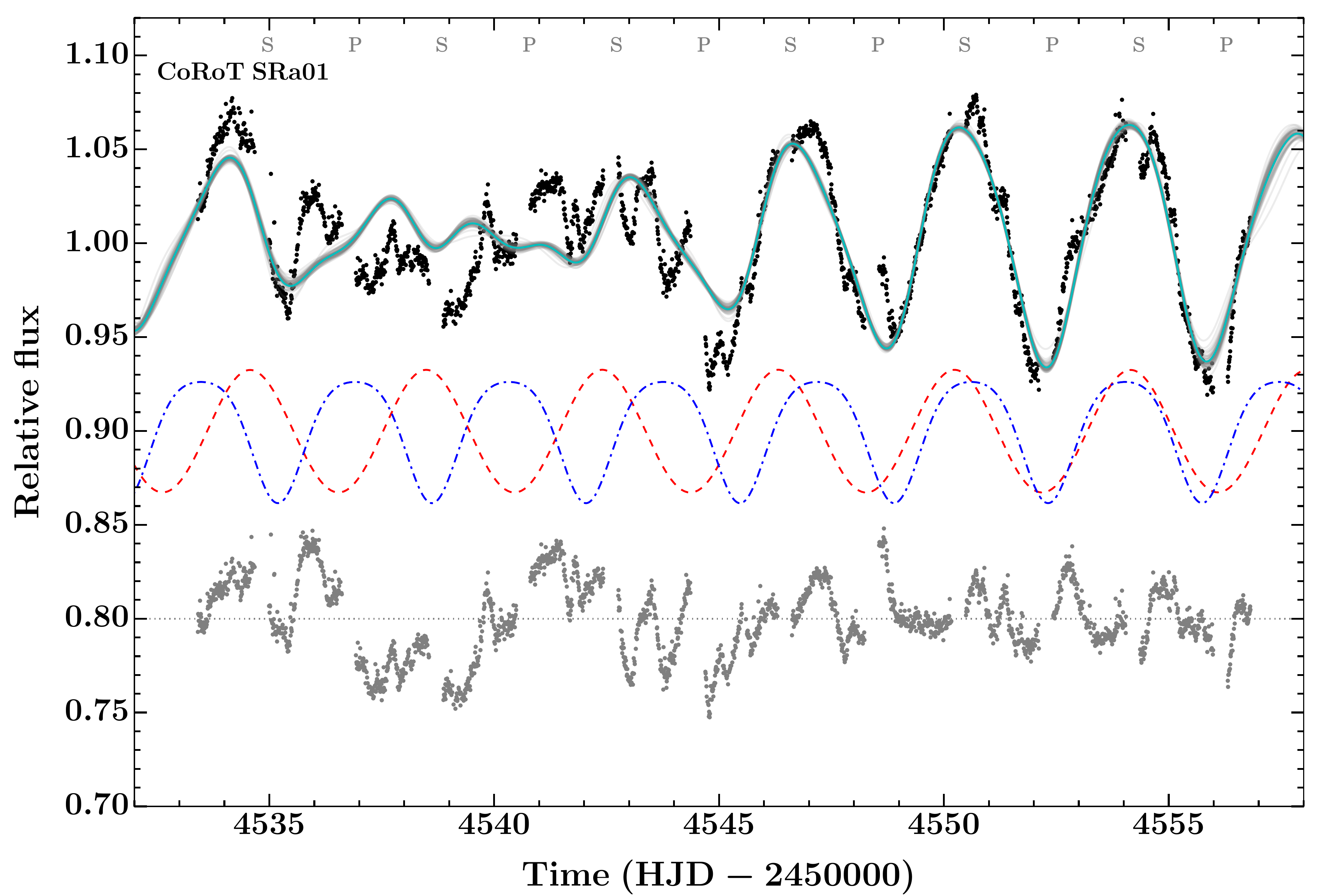}
   \caption{Out-of-eclipse 2008 CoRoT light curve (black) with the two-spot model (cyan) and individual light curve draws from the converged MCMC walkers (grey). Vertically offset are the individual spot models for the primary and secondary stars (red dashed and blue dot-dashed lines, respectively) and the residuals (grey).}
   \label{SRa01_spot_LC}
\end{figure}

\begin{figure}[t!]
  \centering
      \includegraphics[width=\linewidth]{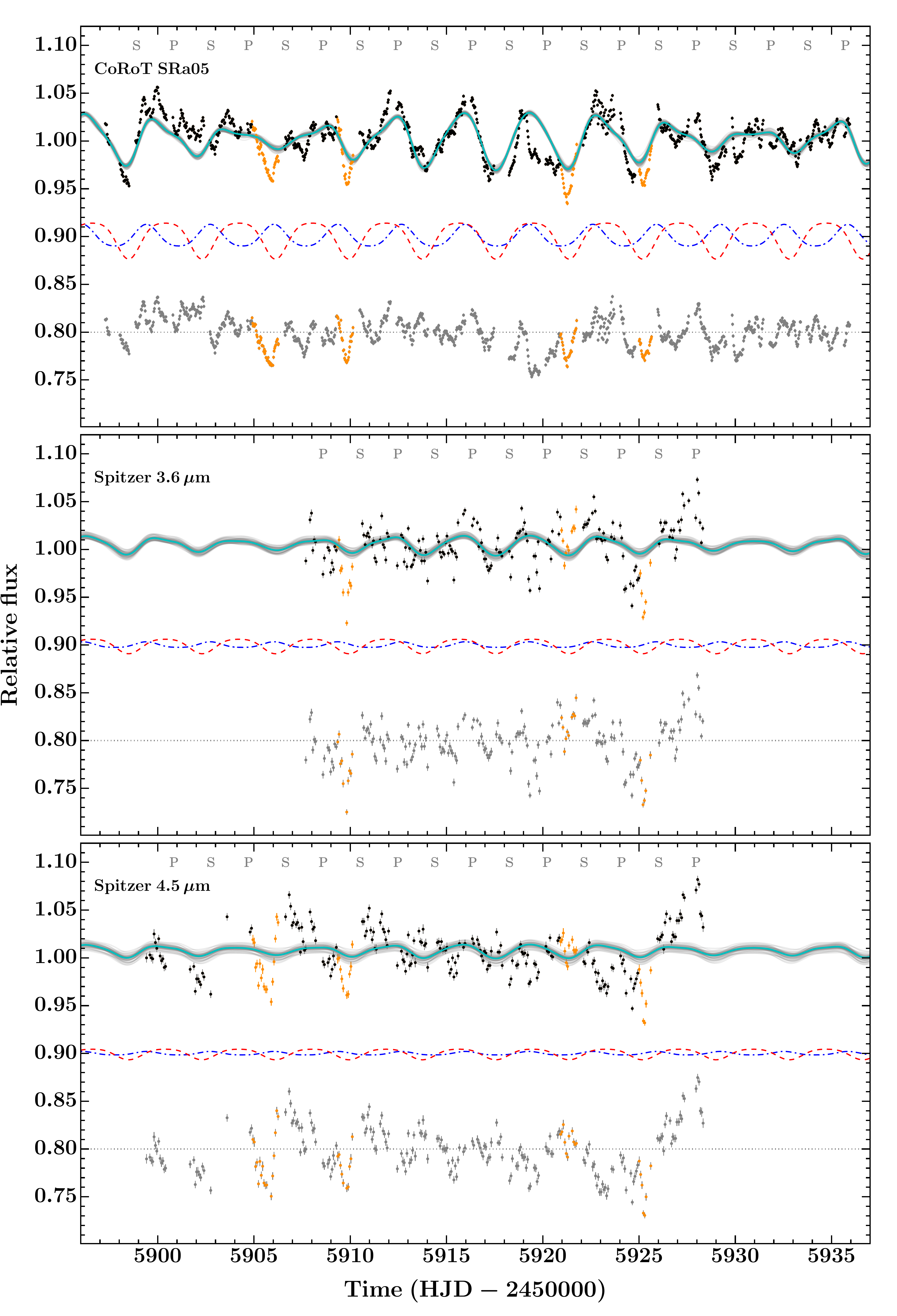}
   \caption{\emph{Top}: out-of-eclipse 2011/2012 CoRoT light curve (black) with the two-spot model (cyan) and individual light curve draws from the converged MCMC walkers (grey). Data masked in in the fit are shown by orange points. Vertically offset are the individual spot models for the primary and secondary stars (red dashed and blue dot-dashed, respectively) and the residuals (grey). \emph{Middle and bottom}: as above for \emph{Spitzer} 3.6 and 4.5 $\mu$m, respectively. All plots share common axes.}
   \label{spot_LCs}
\end{figure}

Figures~\protect\ref{SRa01_spot_LC} and \ref{spot_LCs} show the results of fitting the OOE light curves with the two-spot model (cyan) in each band. Vertically offset are the individual spot models (red dashed and blue dot-dashed for spots on the primary and secondary stars, respectively) and the residuals. The masked data points, which were not used in the fit, are shown in orange. We construct the two-spot model plotted in the figures (cyan) by marginalising over the parameters of the fit, i.e. we take the mean of 200 spot light curves drawn from the converged MCMC walkers (selected individual spot light curves are shown in grey).

The majority of the large scale structure in both CoRoT light curves can be reproduced by the constructive and destructive interference of starspot signals (Fig.~\protect\ref{SRa01_spot_LC} and \ref{spot_LCs}, top plot). There are however differences: in the 2008 CoRoT light curve these are most notable in the first half of the light curve as the fit is driven by the sinusoidal modulations in the second half; some of the differences could therefore result from spot evolution, as this is not incorporated in our model. In the 2011/2012 light curves the most obvious discrepancy is seen around rHJD $\sim$ 5920-5928, but given the amplitude of the flux drop around rHJD $\sim$ 5925 in all three bands, these variations may not be dominated by spot modulation. An alternative explanation could again be spot evolution; the CoRoT light curve spans $\sim$40 days so significant evolution is possible. Indeed, the spot signals in the 2008 and 2011/2012 CoRoT light curves are not in phase, presumably also due to spot evolution. A more flexible method to model the signatures of starspots would be to use Gaussian processes (\citealt{Rasmussen06}; Paper 1) as these would naturally capture any starspot evolution. It is important to note however that such an approach does not yield the physical parameters of the spots themselves, rather it gives an overall representation of their effect on the light curve\footnote{We considered modelling each of the CoRoT and \emph{Spitzer} light curves as the sum of two different Gaussian processes (GPs), which have different covariance properties to disentangle the spot and `non-spot' contributions, but this would not directly yield a solution for each star individually.}.

The parameter values and uncertainties for our simple two-spot models for both the 2008 and 2011/2012 runs are given in Table~\ref{spot_par}. In both cases we find that the solution converges on periods of $\sim$3.8--3.9 days for one spot and $\sim$3.3--3.4 days for the other. Both models also favour one large and one small spot at mid-to-high latitudes.
The spot sizes, and hence overall spot coverages, are consistent with previous photometric studies of T\,Tauri stars \citep[e.g.][]{Grankin98} as well as analyses of optical TiO absorption bands and radius excesses in low-mass stars \citep[e.g][]{ONeal98,ONeal04,Jackson09,Jackson13,Jackson14}.
We also see clear evidence of the well-known degeneracy between the spot size and temperature in the converged MCMC distributions and advise caution when interpreting the values for these parameters: while the quoted values capture something of the magnitude of the uncertainty they do not reflect the complexity of the correlation. We repeated our spot model analysis with different start points for each MCMC `walker' to assess the sensitivity of our solution to our initial positions: the model converges on parameter values comfortably within the uncertainties quoted in Table~\ref{spot_par} in all cases. 
In addition to our spot analysis, we also inspected the power spectra of the CoRoT SRa01 and SRa05 OOE light curves and resolve the two rotation periods in both cases, finding consistent values to our spot model.

Both the 2008 and 2011/2012 models find similar spot periods (3.9 and 3.4 days, and 3.8 and 3.3 days, respectively). If we are sensitive to spots on one star only, this suggests active latitudes that are quasi-stable over many years. However, as both stars are likely heavily spotted we are probably retrieving signatures from both (i.e. each spot in our model attempts to represent the global spot contribution to the light curves from that star). This would suggest slightly different rotation periods for the two stars: one close to or at the binary orbital period (possibly signifying synchronisation) and the other rotating slightly faster. Given the possibility of differential rotation and the simplified nature of our model we refrain from making a more quantitative statement here, but return to this point in section \ref{var_consistency} where we determine the $v \sin i$ of both stars finding consistent rotation periods to our spot model.

To assess the validity of our model we compute a simple statistic, namely the ratio of the root mean square of the residuals to that of the raw data: $\rm{r.m.s(residuals) / r.m.s(raw ~ data)}$. For the 2008 and 2011/2012 CoRoT light curves we find values of 0.54 and 0.69, respectively which, given the simplified nature of the model, validates the underlying prescription, i.e. the large scale structure in the optical CoRoT light curves can arise from the constructive and destructive interference of starspot signals at two slightly different periods. The \emph{Spitzer} light curves, however, are significantly less well fit, having values of 0.94 and 0.97 for the 3.6 and 4.5 $\mu$m bands, respectively. The spot model is not validated in the \emph{Spitzer} bands for three reasons: the two stars only contribute 66 and 55\% of the flux in the 3.6 and 4.5 $\mu$m bands, respectively (the remaining flux comes from the dust in the cavity); starspots are expected to have a reduced effect in the IR; and extra sources of variability arising from processes near, but not at, the stellar surface are expected to significantly affect these bands \citep{Morales-Calderon11,Cody14,Rebull14}.

\begin{table*}[t!] 
\centering 
\caption[]{Parameters of the two-spot model for the 2008 CoRoT light curve (\emph{top}) and the 2011/2012 CoRoT and \emph{Spitzer} observations (\emph{bottom}). } 
\label{spot_par} 
\begin{tabular}{lll llllll} 
\hline 
\hline 
\noalign{\smallskip} 
Parameter   &   Symbol   &   Unit   &   \multicolumn{6}{c}{Value}   \\ 
\noalign{\smallskip} 
\hline 
\noalign{\smallskip} 
\multicolumn{9}{c}{{\it 2008 CoRoT observations }} \\ 
\noalign{\smallskip} 
\noalign{\smallskip} 
  &  &  & \multicolumn{4}{c}{{\it Primary}}  & \multicolumn{2}{c}{{\it Secondary ~~~~ }}   \\ 
\noalign{\smallskip} 
\noalign{\smallskip} 
Radius    &   $\alpha$    &    $\degree$    &    \multicolumn{4}{c}{ $\,\,\,\,93\,^{+20}_{-19}$ }    &    \multicolumn{2}{c}{ $\,\,\,\,40.4\,^{+6.3}_{-2.6}$ ~~~~~~~~~~ }    \\  [1.0ex] 
Latitude    &    $\delta$    &    $\degree$    &    \multicolumn{4}{c}{ $\,\,\,\,83.2\,^{+1.0}_{-2.1}$ }    &    \multicolumn{2}{c}{ $\,\,\,\,63.0\,^{+2.5}_{-4.2}$ ~~~~~~~~~~ }    \\  [1.0ex] 
Longitude    &    $\phi_{0}$    &    $\degree$    &    \multicolumn{4}{c}{ $\,\,\,\,20.8\,^{+3.5}_{-3.0}$ }    &    \multicolumn{2}{c}{ $\,\,\,\,165.2\,^{+2.4}_{-2.2}$ ~~~~~~~~~~ }    \\  [1.0ex] 
Rotational period    &    $P_{\rm{rot}}$    &    days    &    \multicolumn{4}{c}{ $\,\,\,\,3.912\,^{+0.018}_{-0.015}$ }    &    \multicolumn{2}{c}{ $\,\,\,\,3.4249\,^{+0.0060}_{-0.0061}$ ~~~~~~~~~~ }   \\  [1.0ex] 
Temperature    &    $T_{\rm{spot}}$    &    $K$    &    \multicolumn{4}{c}{ $\,\,\,\,2670\,^{+315}_{-260}$ }    &    \multicolumn{2}{c}{ $\,\,\,\,2790\,^{+395}_{-330}$ ~~~~~~~~~~ }    \\  [1.0ex] 
\noalign{\smallskip} 
Maximum flux\,*    &    $F_{\rm{max}}$    &        &    \multicolumn{6}{c}{ $\,\,\,\,1.342\,^{+0.086}_{-0.126}$ }    \\  [1.0ex] 
Jitter term    &    $\sigma$    &        &    \multicolumn{6}{c}{ $\,\,\,\,0.01975\,^{+0.00030}_{-0.00031}$ }    \\  [1.0ex] 
\noalign{\smallskip} 
\noalign{\smallskip} 
\noalign{\smallskip} 
\hline 
\noalign{\smallskip} 
\multicolumn{9}{c}{{\it 2011 / 2012 CoRoT and Spitzer 3.6 \& 4.5 $\mu$m observations }} \\ 
\noalign{\smallskip} 
\noalign{\smallskip} 
  &  &  & \multicolumn{4}{c}{{\it Primary}}  & \multicolumn{2}{c}{{\it Secondary ~~~~ }}   \\ 
\noalign{\smallskip} 
\noalign{\smallskip} 
Radius    &   $\alpha$    &    $\degree$    &    \multicolumn{4}{c}{ $\,\,\,\,26.1\,^{+3.2}_{-2.1}$ }    &    \multicolumn{2}{c}{ $\,\,\,\,122.3\,^{+7.5}_{-20.5}$ ~~~~~~~~~~ }    \\  [1.0ex] 
Latitude    &    $\delta$    &    $\degree$    &    \multicolumn{4}{c}{ $\,\,\,\,69.5\,^{+2.8}_{-2.9}$ }    &    \multicolumn{2}{c}{ $\,\,\,\,73.2\,^{+8.3}_{-6.9}$ ~~~~~~~~~~ }    \\  [1.0ex] 
Longitude    &    $\phi_{0}$    &    $\degree$    &    \multicolumn{4}{c}{ $\,\,\,\,194.9\,^{+4.1}_{-4.1}$ }    &    \multicolumn{2}{c}{ $\,\,\,\,248.9\,^{+7.5}_{-7.5}$ ~~~~~~~~~~ }    \\  [1.0ex] 
Rotational period    &    $P_{\rm{rot}}$    &    days    &    \multicolumn{4}{c}{ $\,\,\,\,3.8139\,^{+0.0080}_{-0.0074}$ }    &    \multicolumn{2}{c}{ $\,\,\,\,3.311\,^{+0.011}_{-0.011}$ ~~~~~~~~~~ }    \\  [1.0ex] 
Temperature    &    $T_{\rm{spot}}$    &    $K$    &    \multicolumn{4}{c}{ $\,\,\,\,2720\,^{+394}_{-290}$ }    &    \multicolumn{2}{c}{ $\,\,\,\,3370\,^{+100}_{-286}$ ~~~~~~~~~~ }    \\  [1.0ex] 
\noalign{\smallskip} \noalign{\smallskip} 
    &    &    &     \multicolumn{2}{c}{ \emph{CoRoT} }    &     \multicolumn{2}{c}{ ~~~~ \emph{3.6\,$\mu$m} }    &     \multicolumn{2}{c}{ \emph{4.5\,$\mu$m} } \\
\noalign{\smallskip} 
Maximum flux\,*    &    $F_{\rm{max}}$    &        &    \multicolumn{2}{c}{ $\,\,\,\,1.105\,^{+0.069}_{-0.034}$ }    &    \multicolumn{2}{c}{ ~~~ $\,\,\,\,1.045\,^{+0.029}_{-0.012}$ }    &    \multicolumn{2}{c}{ $\,\,\,\,1.0416\,^{+0.0235}_{-0.0094}$ }    \\  [1.0ex] 
Jitter term    &    $\sigma$    &        &    \multicolumn{2}{c}{ $\,\,\,\,0.01488\,^{+0.00034}_{-0.00035}$ }    &    \multicolumn{2}{c}{ ~~~ $\,\,\,\,0.0208\,^{+0.0016}_{-0.0014}$ }    &    \multicolumn{2}{c}{ $\,\,\,\,0.0238\,^{+0.0017}_{-0.0016}$ }    \\  [1.0ex] 
\noalign{\smallskip} 
\hline 
\end{tabular} 
\begin{list}{}{} 
\item[*] From an unspotted photosphere. 
\end{list} 
\end{table*}

\subsubsection{Colour curves}

\begin{figure}[t!]
  \centering  
  \includegraphics[width=\linewidth]{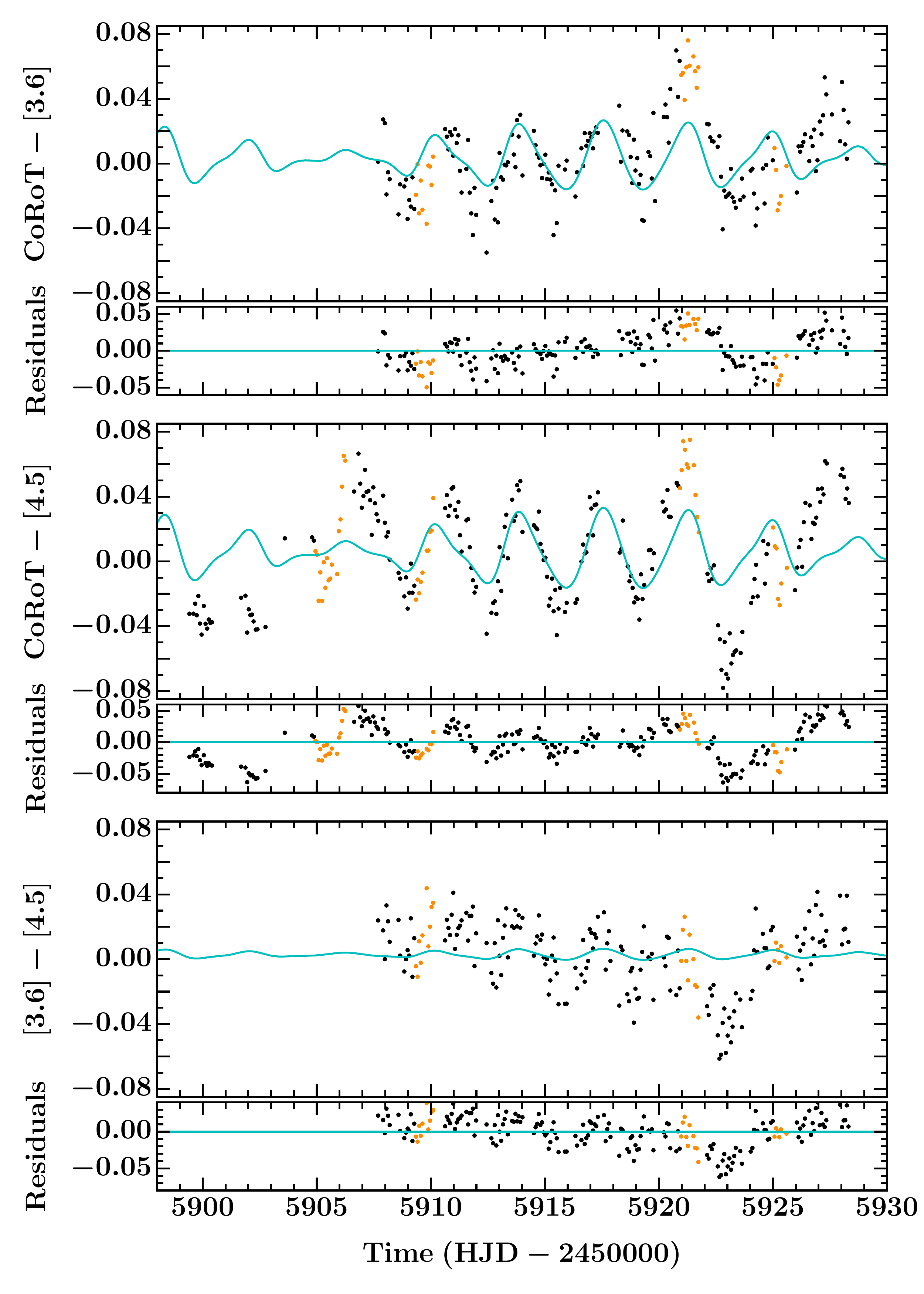}  

   \caption{CoRoT and \emph{Spitzer} colour vs. time relations. \emph{Top-to-Bottom}: CoRoT--[3.6], CoRoT--[4.5] and [3.6]--[4.5], with residuals immediately below in each case. Black points show data used in the spot modelling, orange points indicate data that was masked in the fit and the cyan line represents the colour signature of the spot model. All colour-time relations display quasi-sinusoidal variations, which can most clearly be seen comparing the widely separated bandpasses of the CoRoT-[4.5] colour space. All plots share common axes. }
   \label{col_curves}
\end{figure}

Although the spot models appear to broadly reproduce the structure in the CoRoT light curves they are not able to reproduce the \emph{Spitzer} bands, as expected and as discussed in section~\ref{spot_model}. To confirm the validity of our model we constructed colour vs. time plots (colour curves) to investigate the colour signatures between the different bands in the 2011/2012 dataset. Fig.~\protect\ref{col_curves} shows the three colour curves (CoRoT -- [3.6], CoRoT -- [4.5] and [3.6] -- [4.5]; top to bottom) with the spot model in cyan and residuals immediately below. The colours are calculated from relative magnitudes (rather than relative fluxes as used in the spot modelling) and hence we denote the \emph{Spitzer} 3.6 and 4.5 $\mu$m bands as [3.6] and [4.5], following convention. For both CoRoT--\emph{Spitzer} colour curves, each data point was calculated by binning the CoRoT data in a short time period around each \emph{Spitzer} observation.

The most obvious colour signatures are seen in CoRoT -- [4.5] (middle plot) as they are the most widely separated bandpasses. Between rHJD $\sim$ 5912--5920, when the LAVs dominate the CoRoT light curve, the spot model adequately reproduces the colour signatures (given the simplified nature of the model); this is also true for the corresponding CoRoT -- [3.6] variations. There are two times when the spot model does not reproduce the CoRoT--[4.5] colour variations: a) at rHJD $\sim$ 5905--5907 the colour is significantly redder than the spot model predicts but this is due to an increase in the 4.5 $\mu$m flux, which is not matched in the CoRoT band, i.e. it does not appear to be due to spots, and b) at rHJDs between $\sim$ 5920--5928 when the system becomes first redder and then significantly bluer. The initial reddening is also seen in the CoRoT--[3.6] colour but not the subsequent change towards the blue; the latter appears to be due to a dip in the 4.5 $\mu$m light curve that is only partially reproduced in the CoRoT and 3.6 $\mu$m bands; again clearly not due to spots.

From the residuals of the spot model in both flux and colour spaces it is clear that there is extra variability present that is not well described by starspots, even accounting for their evolution. The question now becomes: what is the physical origin of this additional variability?

\subsection{Colour--magnitude space}

\begin{figure*}[t!]
  \centering  
  \includegraphics[width=0.33\linewidth]{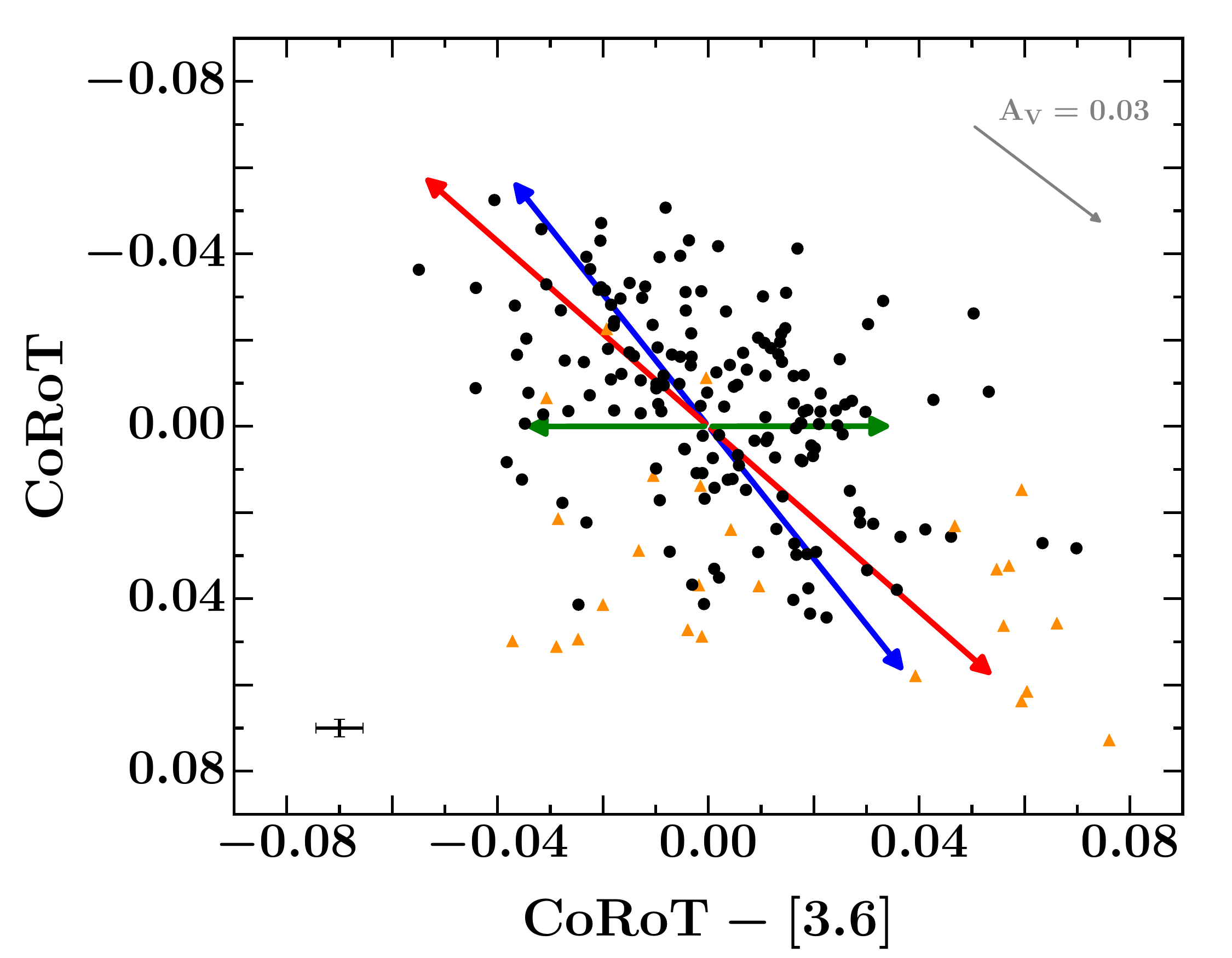}
  \includegraphics[width=0.33\linewidth]{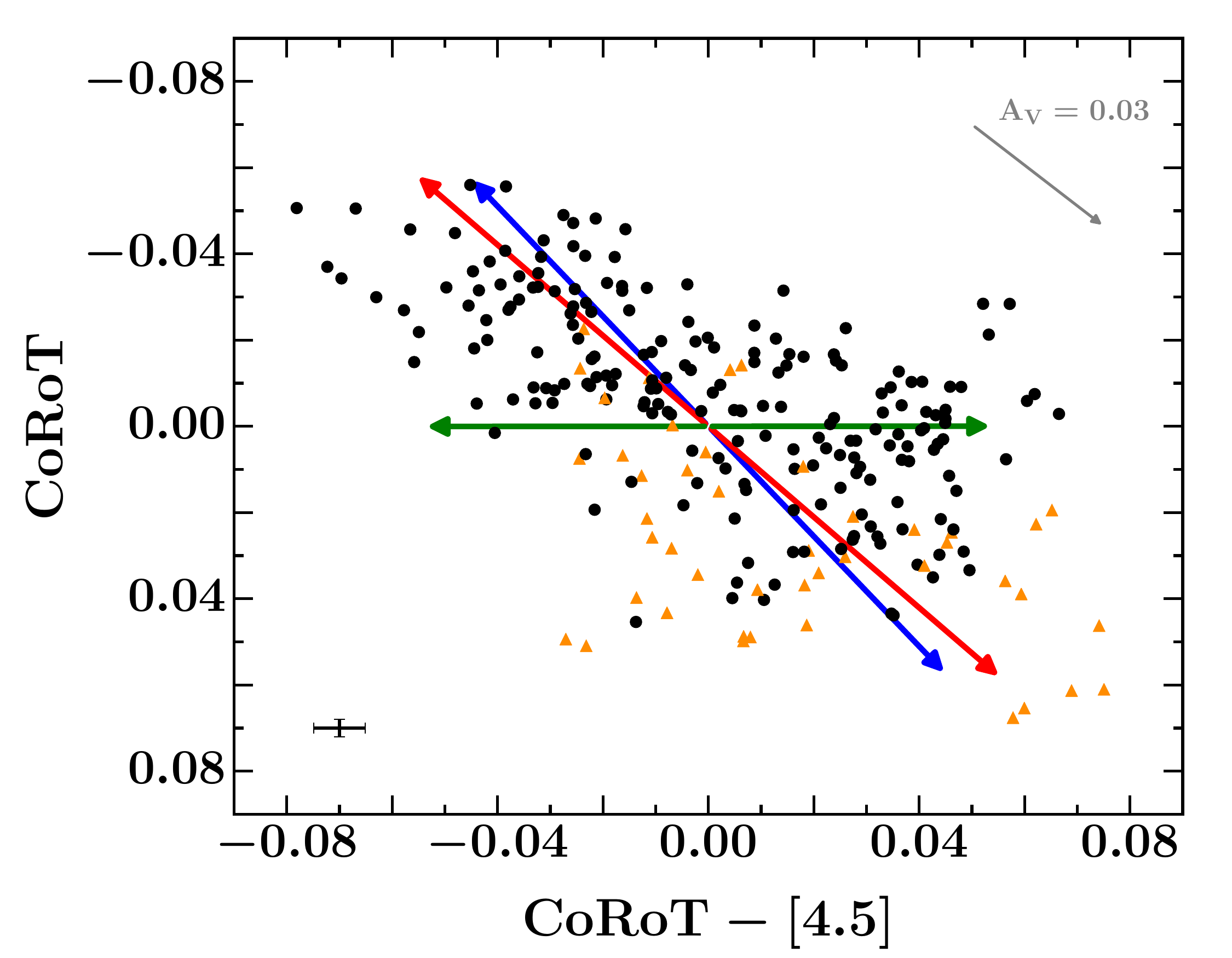}
  \includegraphics[width=0.33\linewidth]{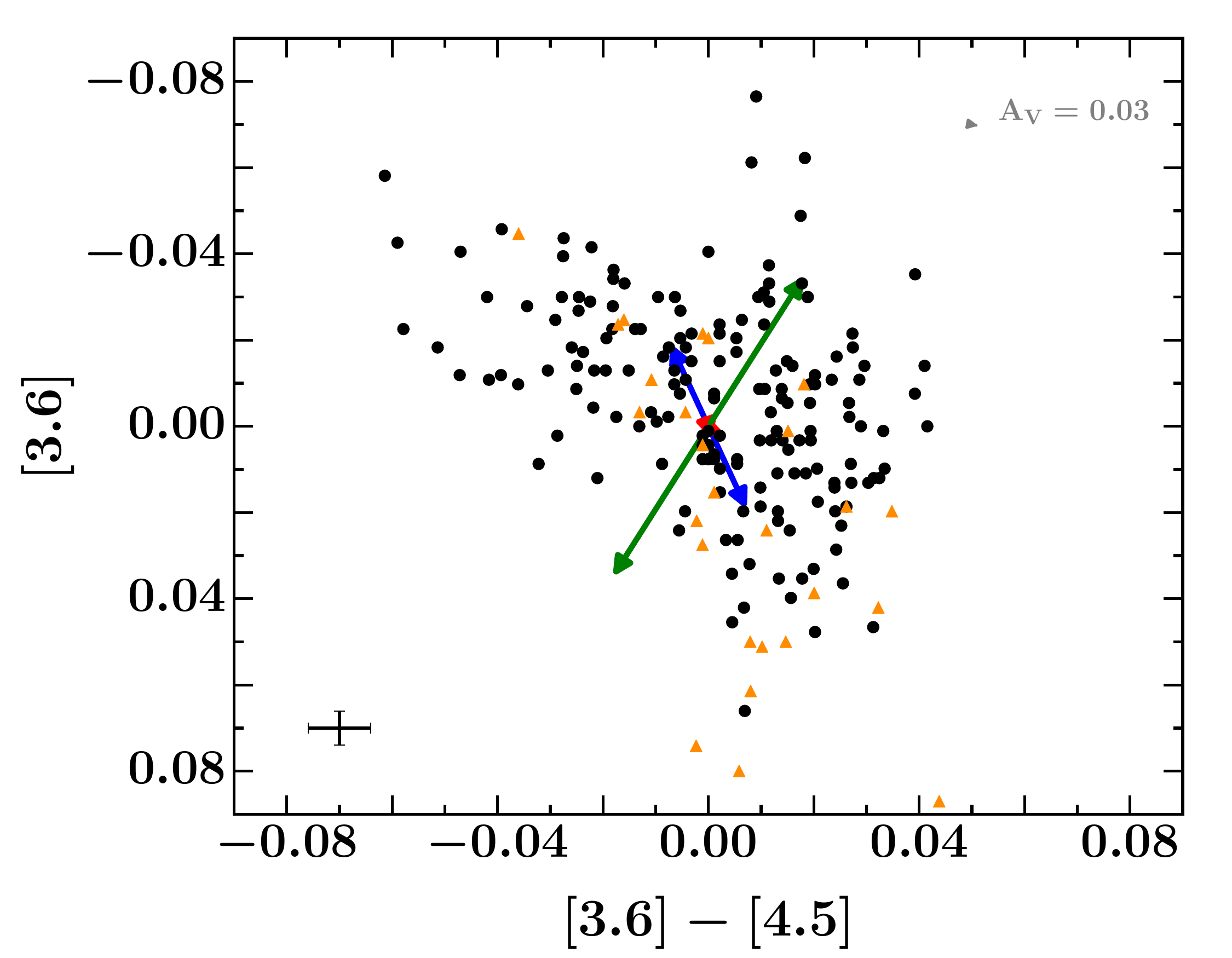}

 \includegraphics[width=0.33\linewidth]{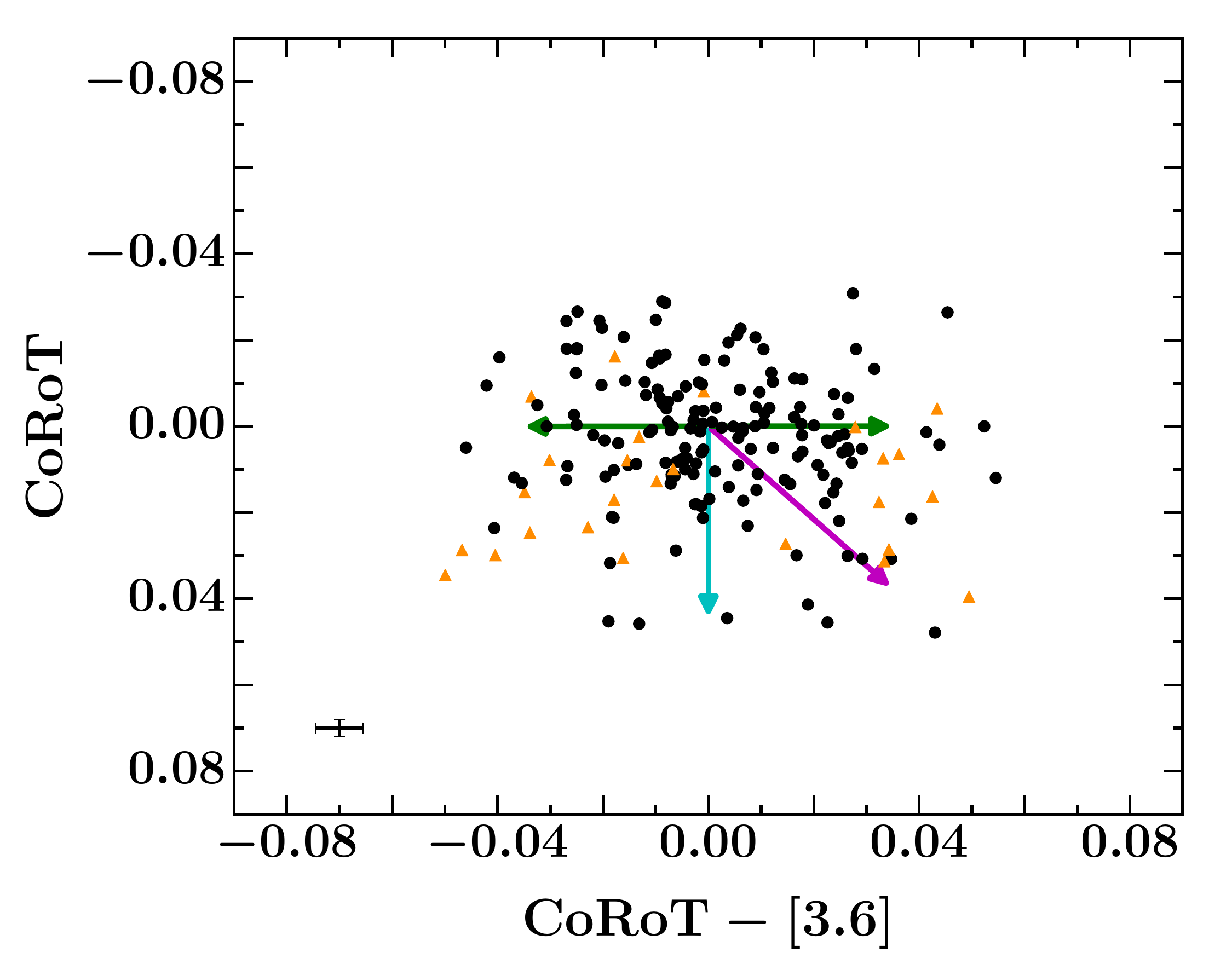}
  \includegraphics[width=0.33\linewidth]{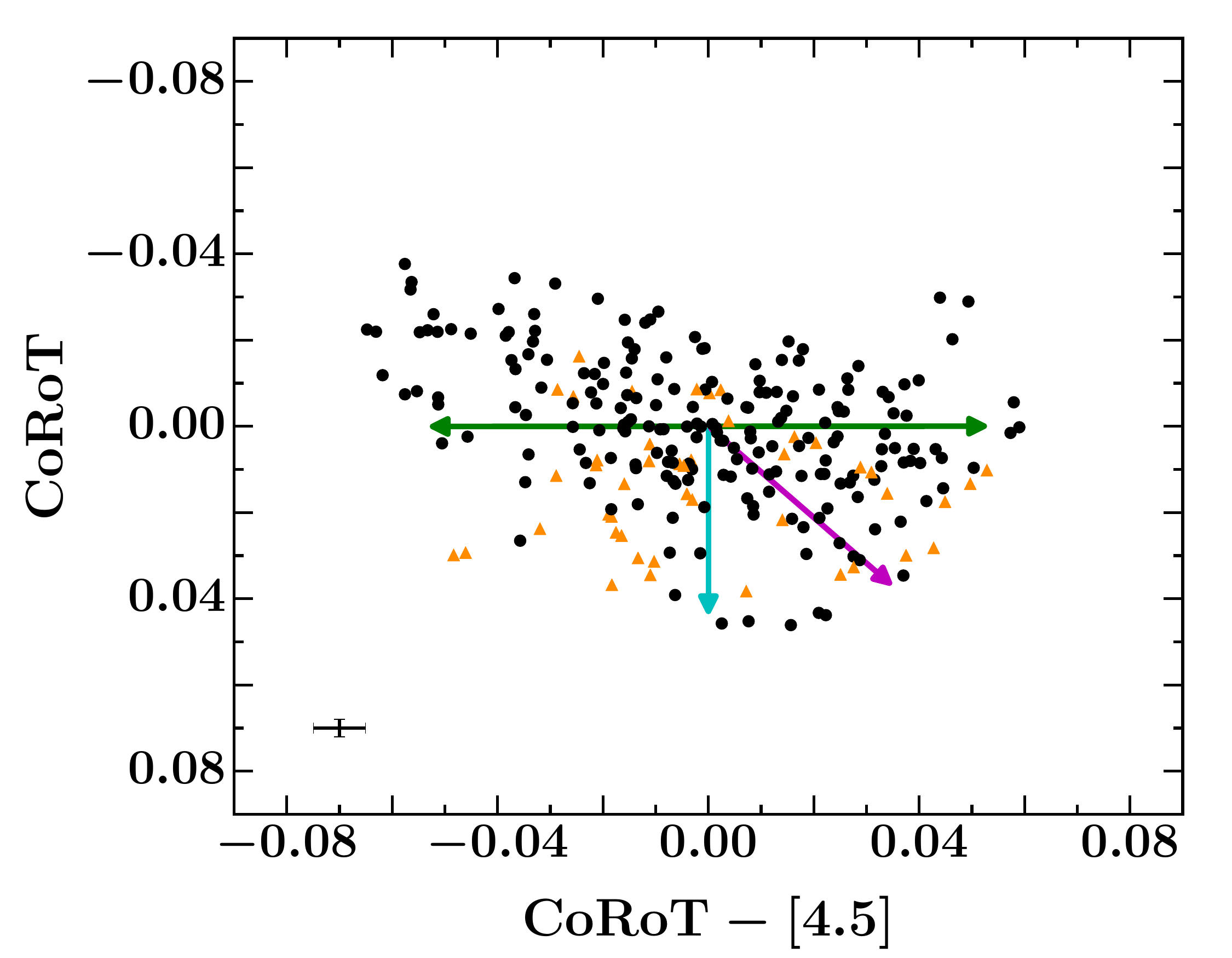}
  \includegraphics[width=0.33\linewidth]{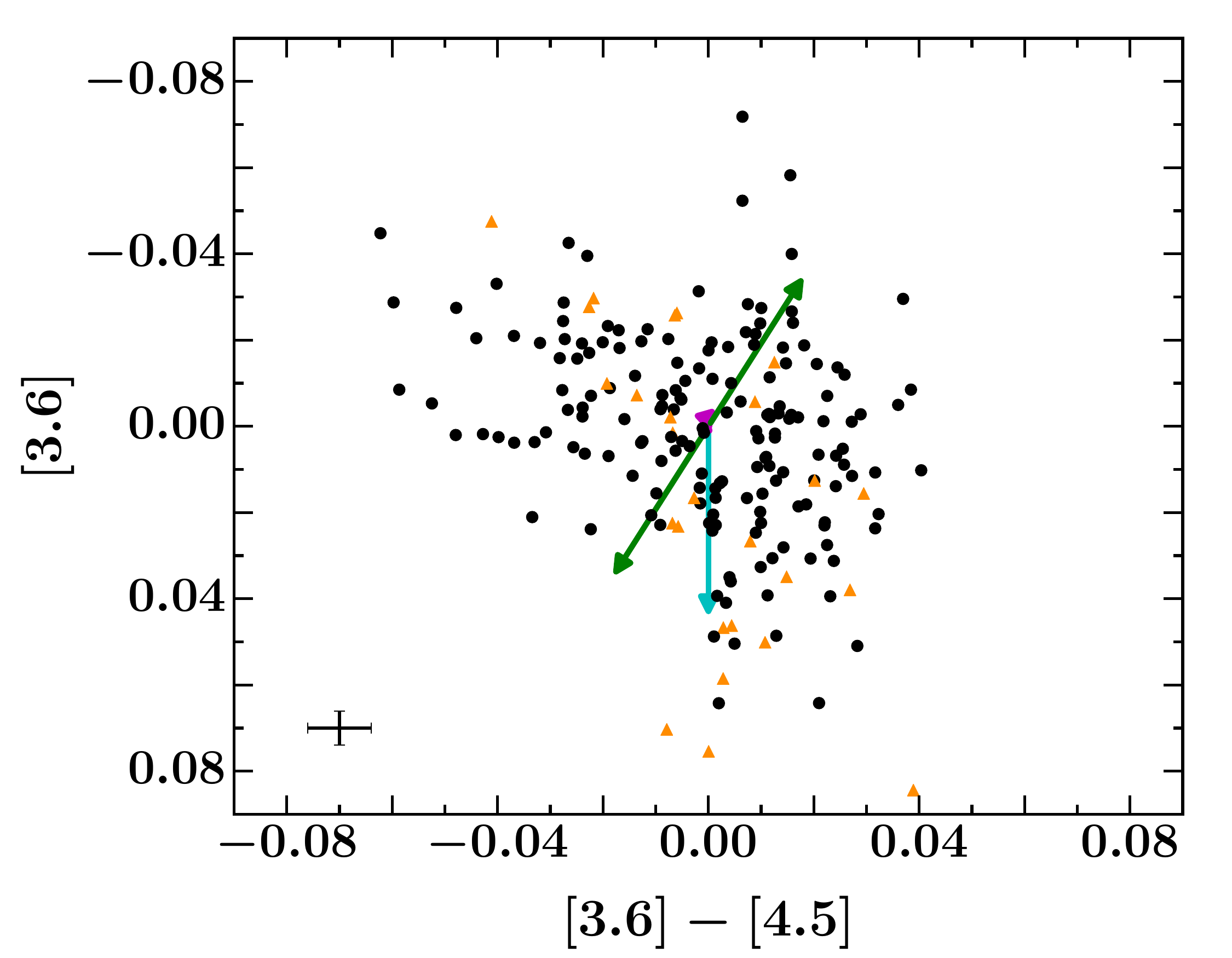}
   \caption{Colour-magnitude diagrams. \emph{Top row, left to right}: raw light curve data (black points) in the CoRoT vs CoRoT--[3.6], CoRoT vs CoRoT--[4.5] and [3.6] vs [3.6]--[4.5] spaces. Data masked in the spot modelling are shown by the orange triangles. Coloured arrows show the effects of different types of variability: cold spots (blue), hot spots (red) and dust emission (green). Grey arrows indicate the direction of interstellar dust extinction. \emph{Bottom row, left to right}: same as for the top row but showing the residuals of the spot modelling. Magenta and cyan arrows indicate the effects of obscuration by small and large dust grains, respectively. The error bar in the bottom left corner of each subplot shows a representative uncertainty for each data point. All plots share common axes.}
   \label{col_mag}
\end{figure*}

Colour-magnitude diagrams remove the temporal element and show how the variations in the three light curves behave as complete sets. Fig.~\protect\ref{col_mag} shows colour-magnitude plots for the different 2011/2012 CoRoT and \emph{Spitzer} combinations both before and after our spot modelling (top and bottom rows, respectively)\footnote{We also constructed magnitude-magnitude plots and colour-colour plots but found the colour-magnitude space to be more informative.}. Black points indicate photometry used in the spot modelling and orange triangles represent the masked regions. The coloured arrows indicate the directions along which the data should lie for different sources of variability. In the top plots, blue, red and green arrows represent starspots (cold spots), hot spots and dust emission, respectively. 

To calculate the cold and hot spot arrows we approximate the two stars with a single star that has luminosity $L_\star=L_1+L_2$, temperature $T_\star$ such that $ \sigma T_\star^4 = L_1/(4 \pi R_1^2) + L_2/(4\pi R_2^2)$ and radius $R_\star$ such that $\sigma T_\star^4 = L_\star/(4 \pi R_\star^2)$. As with the spot modelling, we then estimate the amount of flux in each band as the sum of the stellar and dust components and simulate the effect of cold and hot spots by substituting a fraction of the stellar flux with emission of the relevant photospheric temperature. The size and temperature of these spots were determined by fitting the amplitudes of the LAVs in the 2011/2012 CoRoT light curve (where the stars contribute all the flux). For the cold spot, we select a temperature of $T_{\rm{cs}} = 3000$\,K covering 13\% of the stellar photosphere and for the hot spot, we select a temperature of $T_{\rm{hs}} = 5000$\,K covering 2\% of the stellar photosphere. The exact values are unimportant as long as they are reasonable\footnote{Changing the spot parameters by over 1000K, for example, has little effect on the direction of the arrows.}.

The green arrows were calculated by varying a fraction of the dust required to explain the SED. Variations in this dust emission could result from: variable amounts of dust in the cavity; variable obscuration of dust by one or both stars, or by the inner edge of the circumbinary disk; or even variable accretion, and hence variable amounts of dust at the temperatures required to strongly emit in the 3.6 and 4.5 $\mu$m bands. We require a variable dust fraction of $\lesssim$\,0.25 to fit the amplitude of the variations in the \emph{Spitzer} 4.5 $\mu$m band. This may seem high but it is important to note that this is an upper limit as it is relative to the minimum mass of dust required to fit the SED. If there is more dust in the cavity, the required variable dust fraction would decrease substantially.

The CoRoT vs. CoRoT-[3.6] and CoRoT vs. CoRoT-[4.5] raw colour-magnitude plots (top left and top middle, respectively) show that the amplitude and general direction of the variations (black points) can be explained through either cold or hot spot modulation, but that there is significant scatter above the formal uncertainties (indicated in the bottom left of each panel) that is consistent with variable dust emission. Conversely, in the [3.6] vs [3.6]-[4.5] plane, neither cold/hot spots nor dust emission can be the dominant source of the OOE variations. It is important to note however that these bands are very close in wavelength and hence caution should be exercised when inferring trends, unless they are caused by processes that are sensitive to this small difference, e.g. emitting at a temperature whose black-body peak occurs at a wavelength comparable to 3.6 or 4.5 $\mu$m.

Comparing the top row of plots (raw light curves) to the bottom row (residuals of the spot modelling) for CoRoT vs. CoRoT-[3.6] and CoRoT vs. CoRoT-[4.5], we see that the spread in the black points (those which could be due to spots) in the y-direction has decreased, indicating that the contribution from starspots has been substantially reduced. Note that it will not have been fully removed due to the simplified nature of our model. The colour spread is only slightly reduced, shown by a decrease in r.m.s. scatter of only 16\% for both CoRoT vs. CoRoT-[3.6] and CoRoT vs. CoRoT-[4.5]) suggesting that it is not primarily caused by spots, but it is consistent with variable dust emission. However, there are significant variations from the dust emission trend below the median flux, primarily from the masked data (orange triangles), which correspond to short, sharp flux dips. These variations are not consistent with either starspots (due to the sharpness of their features and their colours) or dust emission, but are more likely to be caused by dust obscuration.

We indicate the direction of small and large dust grain obscuration in the residual plots with the magenta and cyan arrows, respectively. For small dust grains, we assume an interstellar extinction law, following \citet{Schlegel98} and \citet{Indebetouw05}, approximating the CoRoT bandpass with the $r$-band. Obscuration by large dust grains (>\,4.5\,$\mu$m) show grey colour variations and are therefore vertical in all colour-magnitude planes. The amplitudes of the magenta and cyan arrows correspond to obscuring $\sim$10\% of the system flux. The true dust grain size distribution in this system will undoubtedly be complex; here we simply aim to `bracket' plausible directions for dust obscuration effects with the arrows for small and large dust grains. The spread in the residual data below the median is consistent with obscuration by any scale mixture of dust grain sizes.

\subsection{Stability of spots}

\begin{figure*}[]
  \centering
      \includegraphics[width=\linewidth]{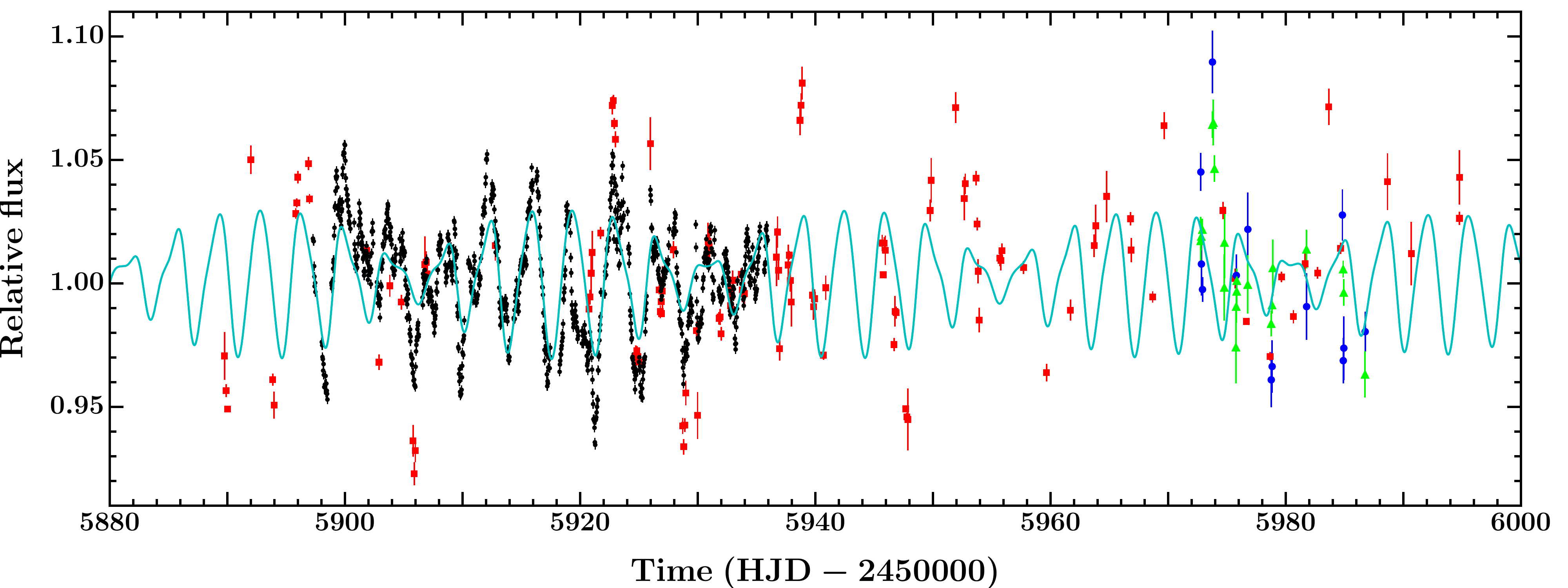}
   \caption{Out-of-eclipse light curves from CoRoT (2011/2012 run), USNO and CFHT ($u$ and $r$-band) spanning 120 days (black points, red squares, blue circles and green triangles, respectively). For clarity, the \emph{Spitzer} light curves are not shown. Our two-spot model, derived from fitting the simultaneous 2011/2012 CoRoT and \emph{Sptizer} light curves, is shown in cyan and extended across the entire period of observations. For clarity, we only show USNO and CFHT data with uncertainties less than 0.015.}
   \label{USNO}
\end{figure*}

To investigate the stability of the out-of-eclipse variability over an extended time period, we compared our spot model derived from the 2011/2012 CoRoT and \emph{Spitzer} observations to data obtained with the USNO 40-inch telescope and CFHT observations (Fig.~\ref{USNO}). The USNO dataset (red squares) spans 106 days (rHJD $\sim$ 5888--5994) and encompasses the 2011/2012 CoRoT observations (black points). To vertically align the USNO and CoRoT datasets we calculated the offset between individual observations taken during the CoRoT run and subtracted off the mean.
The USNO data is a reasonable match to the CoRoT data during the common time period, although there are some discrepancies, e.g. rHJD $\sim$ 5920, where the USNO data appears to more closely follow the \emph{Spitzer} trend (note: the \emph{Spitzer} light curves are not shown for clarity). Beyond the CoRoT dataset the spot model does not appear to closely match the USNO or CFHT (blue and green) observations. This could be due to spot evolution, which is almost certainly present to some degree given the time period covered. We note, however, that the apparent agreement of our model and the USNO and CFHT data is very sensitive to the spot periods: adjusting these slightly, yet remaining within their uncertainties, allows us to obtain a better agreement than shown here but significant discrepancies still exist, which we attribute to spot evolution and additional variability, as previously discussed.

\section{Spectroscopic variability: emission line profiles}
\label{spec_var_sec}

\subsection{H$\alpha$}
\label{Halpha_sec}

In a young stellar system the H$\alpha$ profile can originate from either the star itself, i.e. from magnetically active regions in the chromosphere, or from accretion-related processes. The latter can occur through either simple or complex accretion structures. 

\begin{figure}[t!]
  \centering  
  \includegraphics[width=\linewidth]{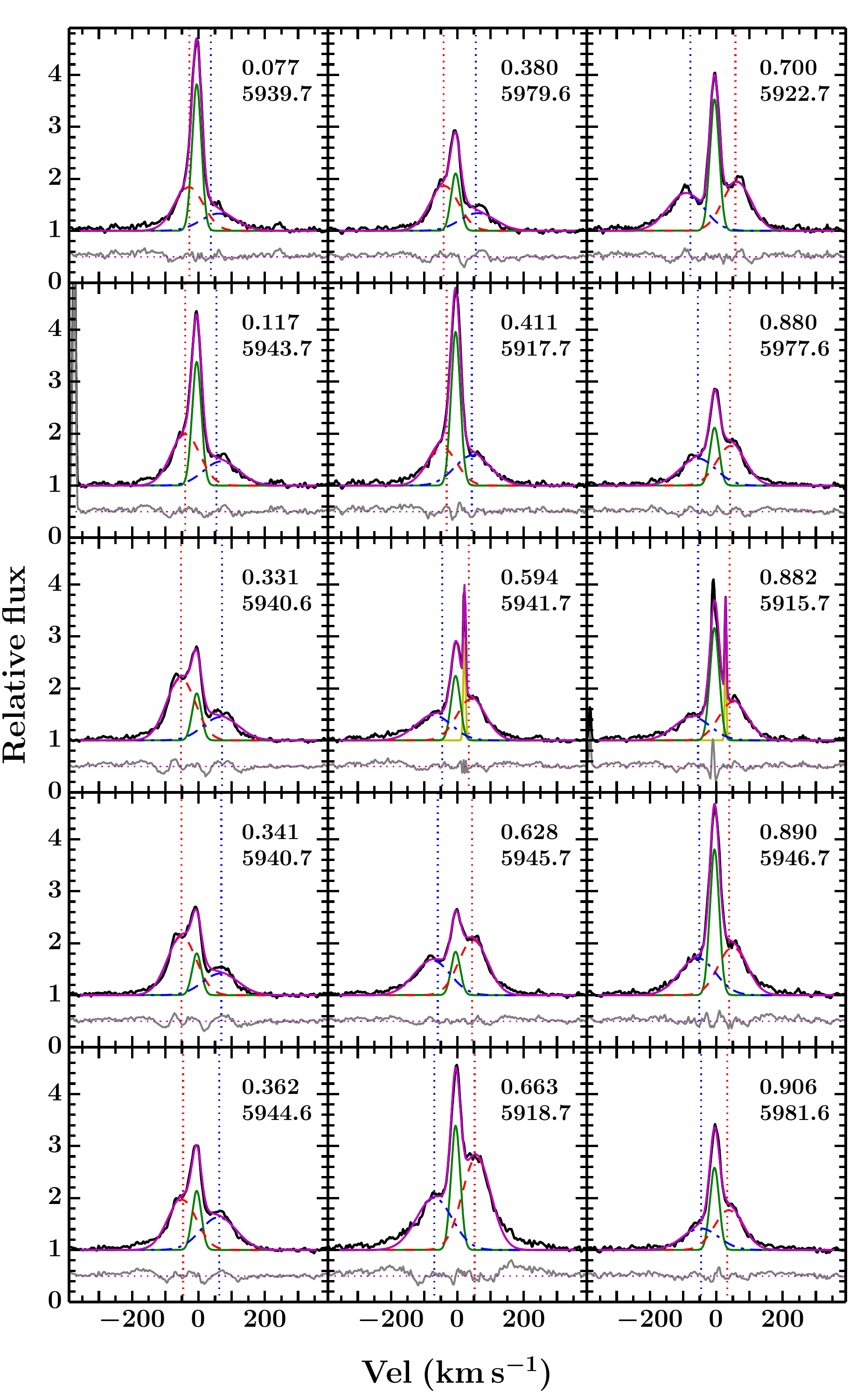}
   \caption{15 VLT/FLAMES H$\alpha$ profiles ordered in phase throughout the binary orbit (phase and rHJD indicated in each subplot). The spectra (black) are modelled as the sum of three Gaussians (magenta). In each case, the central, narrow, static Gaussian (green) is nebula emission and the varying red dashed and blue dot-dashed Gaussians correspond to emission associated with the primary and secondary stars, respectively. The vertical dotted red and blue lines indicate the radial velocities of the primary and secondary stars respectively, at the time of observation. Vertically offset below are the residuals of the model in grey. In two profiles, phases 0.594 and 0.882, additional Gaussians (yellow) are present to account for spikes, which we attribute to cosmic rays.}
   \label{Ha_plot}
\end{figure}

The VLT/FLAMES H$\alpha$ profiles are shown in Fig.~\protect\ref{Ha_plot} (ordered in binary orbital phase; indicated in the top right of each subplot). The H$\alpha$ feature consists of a 3-component emission profile: a central, narrow, static, nebular component, and two components with varying width, velocity and intensity. The velocities of these two components relative to the centre of mass of the system appear to vary in phase with the mean radial velocities of the two stars (indicated by the red and blue dashed lines for the primary and secondary, respectively), but often slightly exceed the latter. 

The simplest explanation for the stellar emission components is chromospheric emission. In previous studies of single T\,Tauri stars where strong accretion signatures are absent, H$\alpha$ emission profiles have been decomposed into narrow and broad Gaussian structures \citep[e.g.][]{Hatzes95,Skelly09}. The narrow component is commonly attributed to the chromosphere but the broad component is the subject of more speculation: \citet{Petrov94} suggest the broader wings could arise from the circumstellar gas environment whereas \citet{Hatzes95} tentatively propose large-scale mass motions or winds in the chromosphere, and both \citet{Jones96} and \citet{Fernandez04} favour `microflaring', i.e. small flares that cannot be individually resolved (although they ruled out the circumstellar gas environment hypothesis due to the absence of IR excesses in their systems).

In CoRoT\,223992193, it is not obvious that the stellar emission components can be decomposed into narrow and broad structures in the majority of profiles. We therefore seek to explain the H$\alpha$ feature as the sum of three Gaussians: one for the central nebula emission (green) and one each for the two stellar components (red dashed and blue dot-dashed for the primary and secondary stars, respectively). We first subtract off a rotationally broadened absorption spectrum for the two stars \citep[e.g.][]{Gray92,Bouvier13b}. We use solar-metallicity PHOENIX model spectra \citep{Husser13}, selecting temperatures of 3700 and 3600 K for the primary and secondary stars, respectively, and a $\log g$ of 4.0 for both (see Paper 1 for the determination of these values). The rotation periods inferred from our spot modelling imply $v \sin i$ values in the range $\sim$14--20\,km\,s$^{-1}$ for the two stars. In addition, the peaks in the cross-correlation functions used to derive the stellar orbits in Paper 1 were not obviously rotationally broadened with respect to the spectrograph's intrinsic resolution ($\sim$17\,km\,s$^{-1}$). We therefore set $v \sin i$ for both the primary and secondary stars to the spectrograph's resolution, i.e. 17\,km\,s$^{-1}$. We note that at the temperatures and surface gravities expected for CoRoT\,223992193, the H$\alpha$ absorption profile is very weak and so subtracting off rotationally broadened absorption lines has very little effect on the shape of the observed profile.

We modelled all H$\alpha$ profiles simultaneously using {\tt emcee}, stepping through the parameter space 20\,000 times with each of 300 `walkers' (the first 10\,000 steps were discarded as `burn in' and parameter distributions derived from the remainder). Modelling all profiles simultaneously allows certain parameters to be jointly fit for: we fix the velocity and width of the nebula component across all profiles as we expect these to be constant but allow the amplitude to vary because it is dependent on the continuum stellar flux level, which changes depending on how accurate the pointing was, i.e. how much of the total system flux reached the detector through the fibre-fed spectrograph. In addition, we fix the widths of the two stellar Gaussians as it is not clear that they vary substantially between profiles (we later relax this assumption). All other parameters were allowed to vary freely. In two profiles (phases 0.594 and 0.882) additional spikes are visible, which we attribute to cosmic rays; these were incorporated into the fit using extra Gaussians (yellow) but are not discussed further. The parameters of the fit are reported in Table~\ref{Halpha_par}.

Within our framework, we find that the model typically converges with half the stellar components best fit by Gaussians centred at velocities close to the stellar RVs and the other half at velocities $\sim$10 and 20 km\,s$^{-1}$ above (i.e. outside the stellar orbits) for the primary and secondary stars, respectively (see columns 6 and 9 of Table~\ref{Halpha_par}). Given our upper limits on the $v \sin i$ of the two stars ($\sim$19 km\,s$^{-1}$; see section \ref{vsini}), this is consistent with the emitting region being located on the surface of each star. Some profiles are very well explained by a simple three Gaussian model, e.g. phases 0.628 and 0.880, and the rest are reasonably well explained but some display evidence of extra emission at higher velocities than the stellar RVs, most notably at phases of 0.362 and 0.663, and occasionally non-Gaussian stellar peaks (e.g. phase = 0.700). We note that the amplitudes of the central stellar emission generally correlate with the strength of the higher velocity emission, e.g. phase = 0.331 and 0.663. Perhaps the simplest interpretation of the higher velocity emission is prominences, i.e. partially-ionised, magnetically-supported plasma structures existing above the chromospheres \citep[e.g.][]{Donati99}.

Given the presence of higher velocity emission, we modelled the profiles again allowing the widths of the stellar Gaussians to vary but constraining their central velocities to the stellar RVs (using Gaussian priors with standard deviations of 1 km\,s$^{-1}$). This avoided some of the secondary emission being fit by the primary Gaussian and vice versa). We then determined estimates of the full widths at 10\% intensity (FW10) for both stellar components and find that they vary between $\sim$150 -- 500 km\,s$^{-1}$. Seven of the 15 profiles are best fit with secondary widths >270 km\,s$^{-1}$, which is generally interpreted as indicating accretion \citep{White03}. The higher velocity emission could therefore be evidence for non-steady, low-level accretion, which might be expected given the presence of dust in the central cavity and the short-lived $u$-band excess in the CFHT light curves. It is important to note, however, that chromospheric emission would still be present in such a scenario and so each stellar component should be modelled with a narrow and broad Gaussian, rather than a single Gaussian only. The quoted FW10s therefore are not representative of any individual component but rather the general contribution from each star to the composite profile. We opted not to model the H$\alpha$ profiles with a 5-Gaussian model as any results would still be ambiguous. We investigate the relationship between FW10 and binary orbital phase, sometimes seen in other young accreting binary systems, in section \ref{phase_dep}.

In some single T\,Tauri systems, such as  AA\,Tau \citep{Bouvier07}, the H$\alpha$ profiles can be decomposed into simple Gaussian components representing different parts of the accretion flow (hot spots, accretion columns and winds). In others, such as V2129 Oph \citep{Alencar12}, this is not possible but the profiles can be explained by emission from accreting material in the stellar magnetospheres, whose variability is driven by the rotational modulation of non-axisymmetric multipolar components of the stellar magnetic fields. In this latter scenario, the observed stellar emission profiles could be explained using radiative transfer models based on the accretion flow structure generated through three-dimensional magnetohydrodynamical (3D MHD) simulations. Such models could explain both the higher velocity emission and the variable peak shapes. However, to draw meaningful conclusions from this type of analysis one needs a priori knowledge of the stellar magnetic field configurations and strengths, which we do not have. We therefore see this as a potential option for future work. Given the complexity of the problem one would also certainly need multiple observational constraints, i.e. multiple emission lines observed simultaneously (particularly from different series of the same element, sharing a common upper level) to break the degeneracy between chromospheric and different accretion signatures.

We conclude that the majority of the H$\alpha$ profiles are consistent with chromospheric emission. Higher velocity emission is sometimes evident, which could be due to prominences above the stellar chromospheres or, given the velocities involved, could alternatively indicate low-level, non-steady accretion. Accreting binaries typically show H$\alpha$ profiles that are indistinguishable from those of single stars (see e.g. DQ Tau, \citealt{Basri97}; AK Sco, \citealt{Alencar03}). By contrast, the well-separated components seen in CoRoT\,223992193 could enable us to probe accretion/outflow on each star separately. However, we need higher spectral resolution data to investigate the details of the H$\alpha$ profiles and make robust statements.


\subsection{Activity correction to the estimated stellar radii and temperatures}

In Paper 1, we modelled the light and radial velocity curves of CoRoT\,223992193 to obtain radii of $R_{\rm pri}$ = 1.30$\pm$0.04 $R_{\odot}$ and $R_{\rm sec}$ = 1.11$^{+0.04}_{-0.05}$ $R_{\odot}$, along with corresponding effective temperature estimates of $T_{\rm pri}$ $\sim$ 3700K and $T_{\rm sec}$ $\sim$ 3600 K from the system's SED. Given that photometric and spectroscopic activity is seen in CoRoT\,223992193, it is possible that the radii are inflated and the effective temperatures suppressed. 
\citet{Stassun12} provide an empirical correction for activity effects on the radii and temperatures of low-mass stars using H$\alpha$ luminosity and equivalent width. 
Using our estimate of the (chromospheric) H$\alpha$ emission for each star, the empirical Stassun et al. correction predicts that the radii of the primary and secondary stars are inflated by 9.3$\pm$3.8 and 8.3$\pm$3.6\,\%, and the temperatures suppressed by 4.3$\pm$1.5\ and 3.9$\pm$1.4\,\%, respectively.  Applying these corrections would give activity-corrected radii of $R_{\rm pri}$ = 1.19$\pm$0.05 $R_{\odot}$ and $R_{\rm sec}$ = 1.02$\pm$0.06 $R_{\odot}$ and temperatures of $T_{\rm pri}$ $\sim$ 3850K and $T_{\rm sec}$ $\sim$ 3750 K.

\subsection{Search for other emission lines}
\label{other_em_lines}

\begin{figure*}[t!]
  \centering  
  \includegraphics[width=0.9\linewidth]{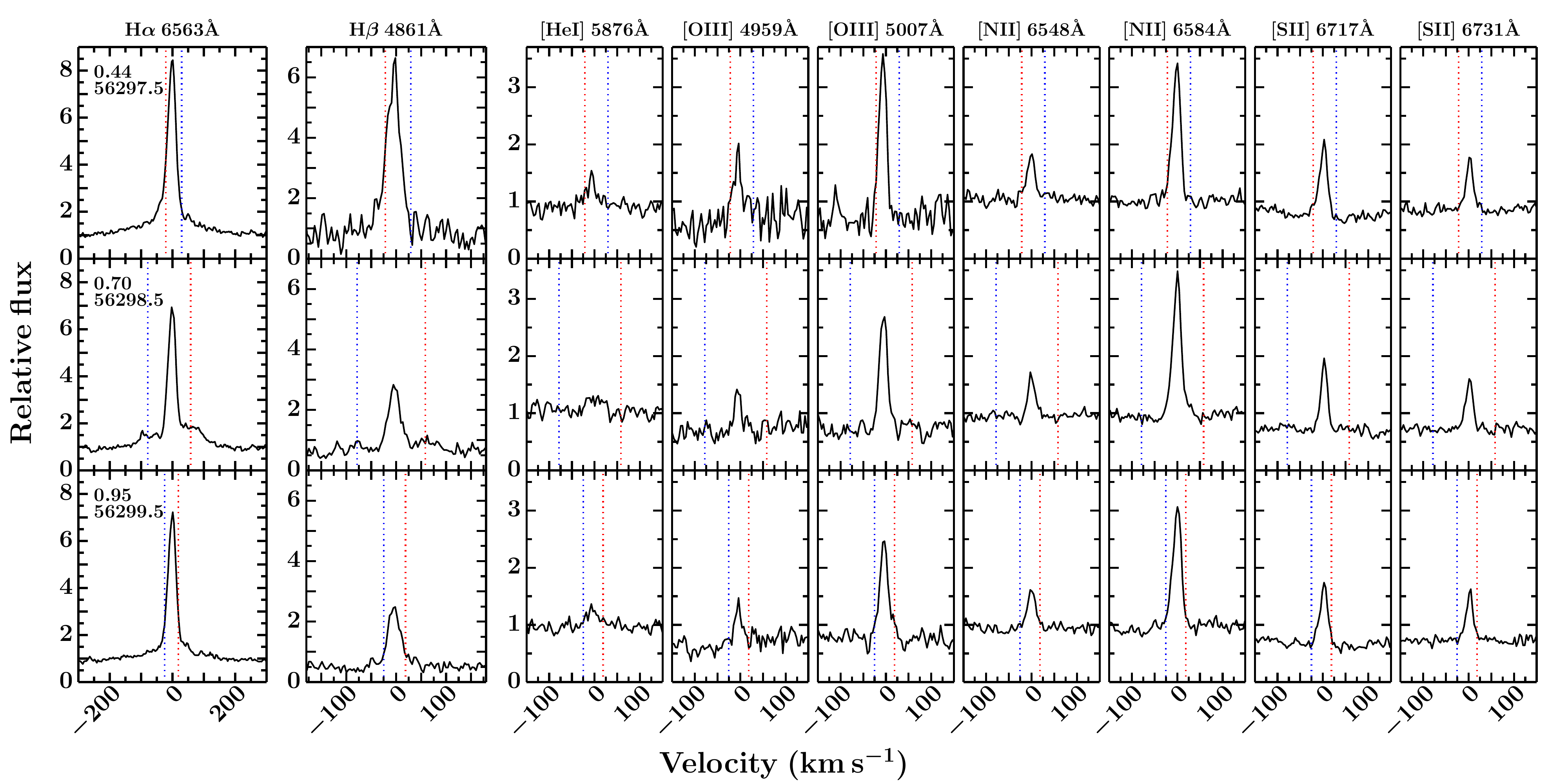}
   \caption{Emission line profiles observed with NOT/FIES. \emph{Left to right}: the permitted emission lines H$\alpha$, H$\beta$ and HeI, and the forbidden emission doublets OIII, NII and SII. \emph{Top to bottom}: the three consecutive nights of observations.The phase and time of each observation are given in the H$\alpha$ panels.}
   \label{FIES_em_lines}
\end{figure*}

In order to obtain more information on the origins of the H$\alpha$ emitting material we searched for emission lines within the wavelength range, 3630 -- 7170 \AA, covered by the three FIES spectra. These lines were H$\alpha$, H$\beta$ (4681\,\AA), HeI (5876\,\AA), and the forbidden emission lines OIII (4959\,\AA\ and 5007\,\AA), OI (6300\,\AA\ and 6364\,\AA), NII (6548\,\AA\ and 6584\,\AA) and SII (6717\,\AA\ and 6731\,\AA). The emission line profiles are shown in Fig.~\protect\ref{FIES_em_lines}\footnote{The OI (6300\,\AA\ and 6364\,\AA) emission feature is dominated by Earth airglow and is therefore not shown.}. We note that the structure of the H$\alpha$ profile matches the FLAMES spectra. We find tentative evidence for H$\beta$ emission associated with the primary star, most clearly seen on the middle night (5$^{\rm{th}}$ January). Unfortunately, the low S/N of the FIES spectra at 4860\,\AA\ prevented us from performing a thorough analysis of the H$\beta$ line profile. Higher S/N spectra are needed to make meaningful statements about the system's H$\beta$ profile. We did not find evidence for emission associated with either star in any of the forbidden emission lines; we see only nebula emission.

\section{Photometric and spectroscopic consistency}
\label{var_consistency}

\subsection{$v\sin i$ estimation}
\label{vsini}

Assuming our spot periods represent the rotation periods of the two stars implies rotational velocities of order $\sim$14--20 km\,s$^{-1}$ for the two stars. The FLAMES resolution is $\sim$17 km\,s$^{-1}$, which is unlikely to be able to robustly resolve the rotational velocities of the two stars. However, FIES has a resolving power of $\sim$12 km\,s$^{-1}$, which is sufficient. The $v\sin i$ of both stars were determined from the three FIES spectra by cross-correlating a slowly rotating M2 spectral standard star (observed with FIES in the same setup) with each of the observed spectra for different rotational broadenings. Analysis of the height and contrast of the CCF stellar peaks implies the same $v\sin i$ for both stars of between 17--19 km\,s$^{-1}$. Taking account of the different stellar radii, this corresponds to rotation periods of 3.5--3.9 days for the primary star and 3.0--3.3 days for the secondary star, which given the $v \sin i$ uncertainties and limitations of the spot modelling, is consistent with the spot-derived periods. The primary star rotation is consistent with synchronisation and the secondary star rotation is slightly supersynchronous.

\subsection{Seeking evidence of spot modulation in VLT/FLAMES spectra}

Four of the VLT/FLAMES spectra were taken simultaneous with the CoRoT/\emph{Spitzer} observations. We sought additional evidence for the spot origin of the large scale optical photometric variability through a bisector span analysis of the Li\,6708\,\AA\ absorption lines in the four simultaneous spectra. Assuming our spot periods do represent the rotation periods of the two stars, and given the combination of the FLAMES resolution being $\sim$17 km\,s$^{-1}$ and the spectra having S/N $\sim$22--26, we did not expect to see unambiguous signs of the spots in the bisector spans, and indeed we do not. This does not mean that spots are not responsible for the large scale variations but simply that the FLAMES data do not possess the resolution and S/N required to unambiguously prove their presence.

\subsection{Correlation between the photometric and spectroscopic variability}

There does not appear to be a correlation between the photometric and spectroscopic variability. Four H$\alpha$ profiles taken at rHJD (phase) = 5915.7 (0.882), 5917.7 (0.411), 5918.7 (0.663) and 5922.7 (0.700) were simultaneous with the 2011/2012 CoRoT and \emph{Spitzer} photometry. The strength of the stellar component of the H$\alpha$ emission does not appear to correspond to either optical or IR continuum flux levels, or variations. 
This lack of correlation does not appear to be an artefact, e.g. the profile at phase = 0.882 shows the smallest stellar emission amplitudes but was taken at the highest S/N (26 as opposed to 22 for the rest).

\subsection{Correlation between binary orbital phase and the spot-corrected photometric and spectroscopic H$\alpha$ variations}
\label{phase_dep}

In other young binary systems that actively accrete from circumbinary disks, the accretion-related photometric and spectroscopic line variations phase with the binary orbit, being typically brightest/strongest around periastron passage in eccentric systems (e.g. DQ Tau: \citealt{Mathieu97,Basri97} and UZ Tau E: \citealt{Jensen07}). Given the circular orbit of CoRoT\,223992193, and theoretical predictions that the strength of phase-dependent accretion-related variations should diminish with decreasing eccentricity \citep[e.g.][]{Artymowicz96}, it is not clear that we should see strong phase-dependent variations in this system. However, more recent 3D magnetohydrodynamic (MHD) simulations show that an eccentric inner disk can exist around an equal mass binary on a circular orbit and that the accretion streams are not stable structures \citep{Shi12}. We therefore investigated the dependence of a) the residuals of our spot model and b) the stellar H$\alpha$ emission profiles on the binary orbital phase.

\begin{figure}[t!]
  \centering  
     \includegraphics[width=\linewidth]{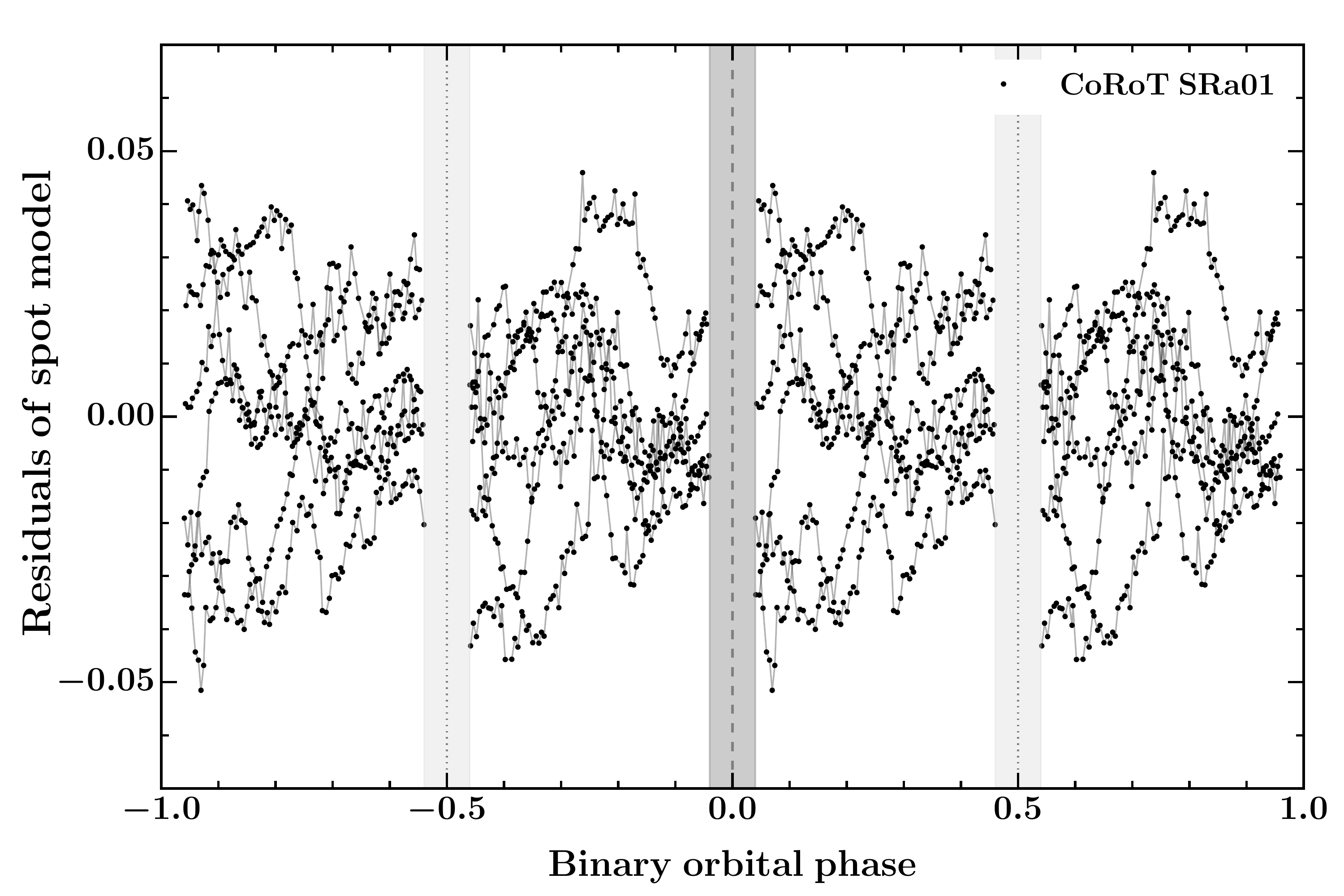}
    \includegraphics[width=\linewidth]{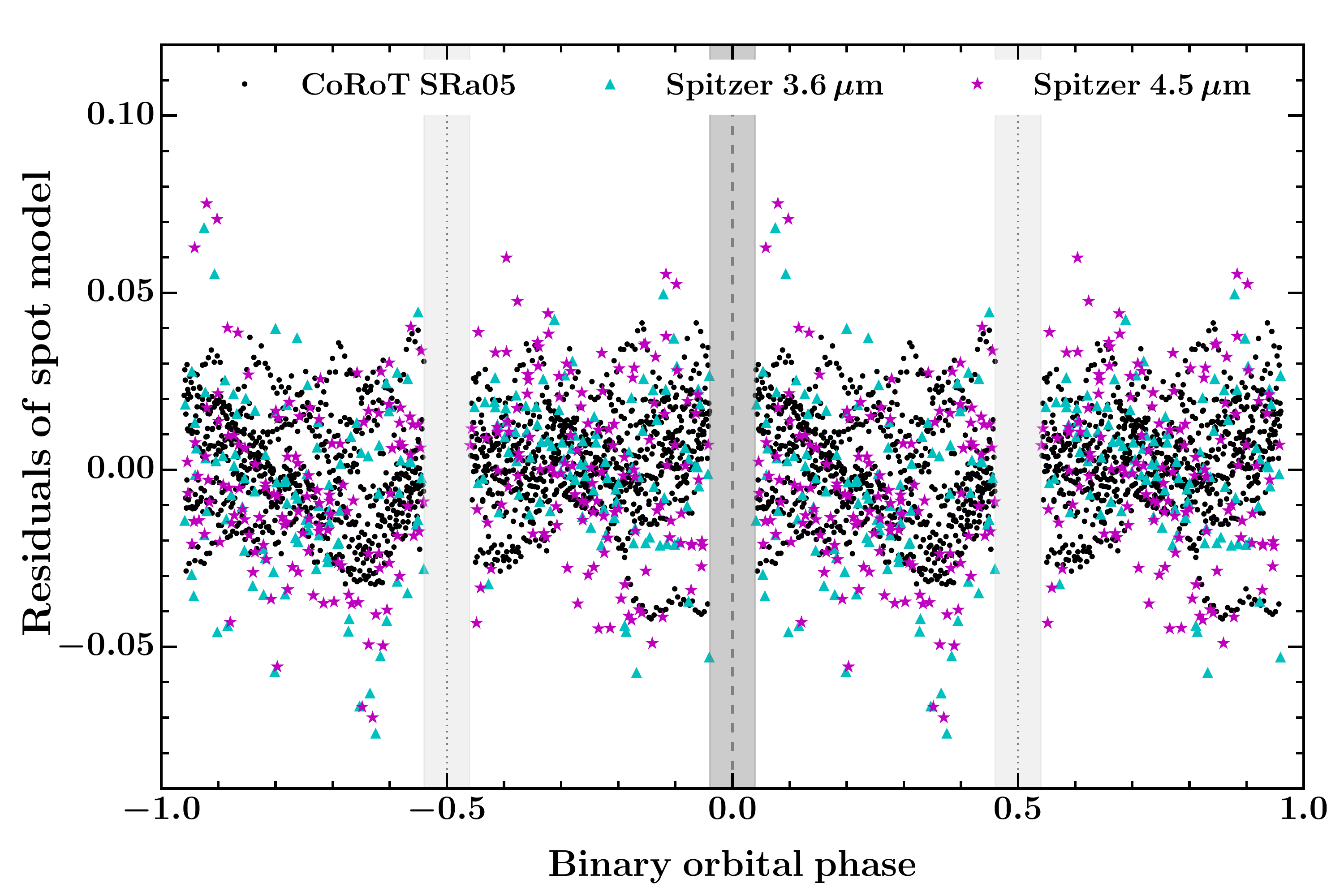}
      \includegraphics[width=\linewidth]{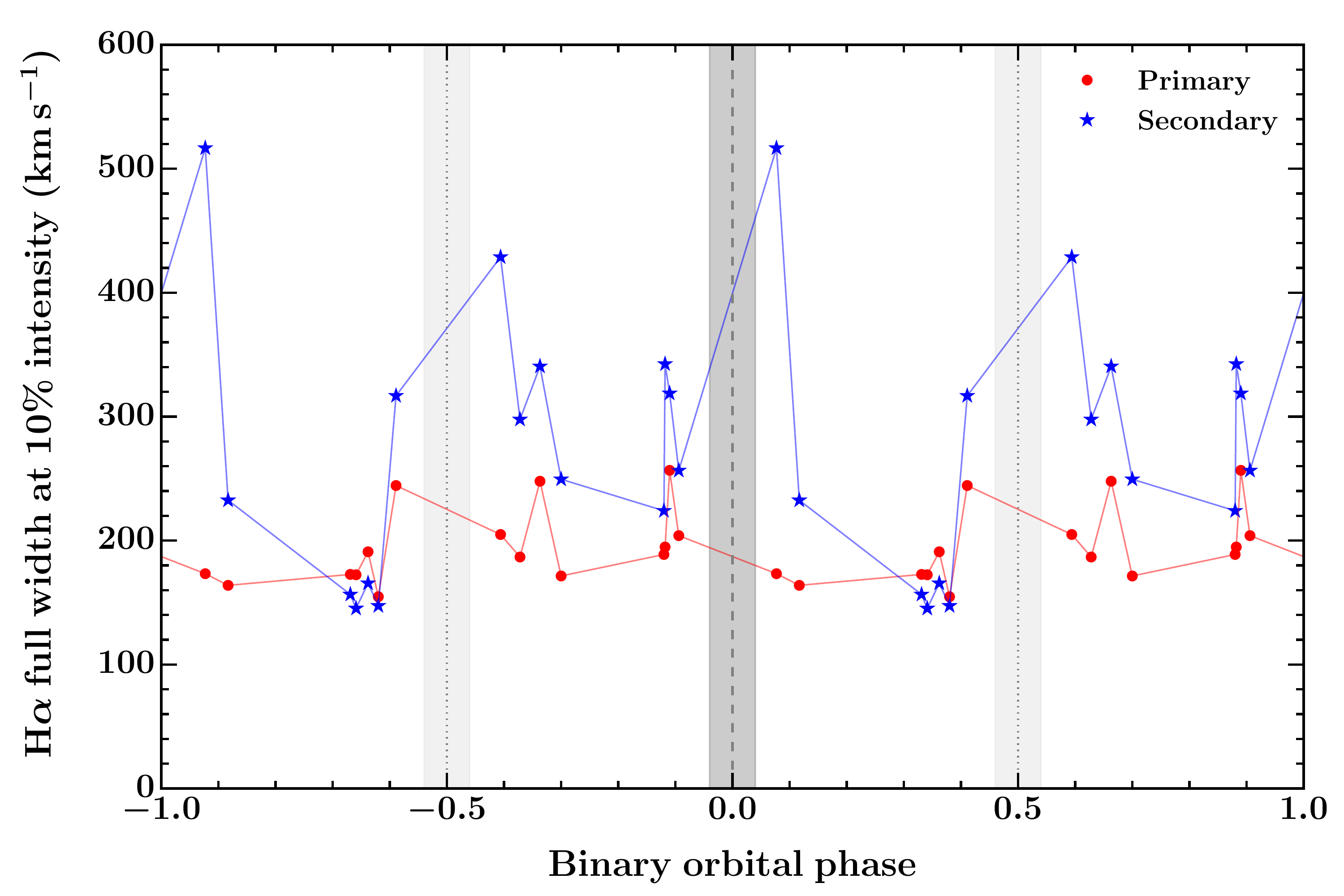}
   \caption{Correlation between binary orbital phase and the spot-corrected photometric, and spectroscopic H$\alpha$, variations. \emph{Top}: residuals of the 2008 spot model folded at the binary orbital period. The dark and light vertical grey bands around phases 0.0 and $\pm$0.5 indicate the phase and duration of primary and secondary eclipses, respectively. \emph{Middle}: same as \emph{top} but for the 2011/2012 CoRoT and \emph{Spitzer} residuals (black points, CoRoT; cyan triangles, \emph{Spitzer} 3.6\,$\mu$m; and magenta stars, \emph{Spitzer} 4.5\,$\mu$m). \emph{Bottom}: Full width at 10\% intensity of the primary and secondary H$\alpha$ emission components folded at the binary orbital period (red points and blue stars, respectively). }
   \label{phot_spec_phase}
\end{figure}

The top and middle panels of Figure \ref{phot_spec_phase} show the phase-folded residuals of the 2008 CoRoT and the 2011/2012 CoRoT and \emph{Spitzer} spot models, respectively. The 2008 residuals do not display an obvious dependence on the binary orbit. However, the 2011/2012 residuals do appear to show a tentative phase-dependence, with peaks in flux occurring around primary and secondary eclipses (phases 0 and $\pm$0.5, respectively). These are perhaps most clearly seen in the \emph{Spitzer} bands, as might be expected given their redder bandpasses. Inspection of the 2011/2012 time-series suggests that the observed peaks around primary and secondary eclipse are driven by variations seen in only a few orbits: if this apparent phase-dependence is due to accretion-related processes, they are non-steady and low-level. Given this, and the simplified nature of our spot model, it is difficult to be conclusive without a longer temporal baseline to assess the significance of the phasing of the observed peaks. Another structure that is apparent in the 2011/2012 phase-folded residuals is the aforementioned short-duration flux dip preceding secondary eclipse (phase$\sim$0.4), which is visible in all three bands. 

We also investigated the phase-dependence of the FLAMES H$\alpha$ profiles. We calculated the full width at 10\% intensity (FW10) of the stellar emission components from the model described in section \ref{Halpha_sec} where the widths of the stellar Gaussians were allowed to vary. The bottom panel of Fig.~\ref{phot_spec_phase} shows the primary and secondary FW10s as a function of binary orbital phase. The FW10 of the primary appears relatively constant at $\sim$200 km\,s$^{-1}$ while the secondary star shows significantly more variation, ranging from $\sim$200-350 km\,s$^{-1}$ with peaks of $\sim$500 and 400 km\,s$^{-1}$ close to primary and secondary eclipses, respectively. It is encouraging that there appears to be a similar structure between the residual photometric 2011/2012 variations in the \emph{Spitzer} bands and the secondary star H$\alpha$ profiles: i.e. peaks in brightness and increases in FW10 around primary and secondary eclipses\footnote{We note that the 2011/2012 CoRoT and \emph{Spitzer} light curves, and the FLAMES H$\alpha$ observations were taken contemporaneously.}. However, two points are worth noting: i) as previously mentioned, the FW10s are calculated from a single Gaussian fit to each stellar emission profile whereas, if we are detecting evidence of non-steady, low-level accretion in the FW10 changes, each stellar profile should more realistically be modelled with both a narrow and broad Gaussian, as chromospheric emission will still be present\footnote{This possibly explains why the profile at phase 0.663 do not appear to show a large FW10: the fit is dominated by the strong, central chromospheric component of the profile.}; and ii) the uncertainty on the FW10s is expected to increase around primary and secondary eclipses due to the stellar components being closer in velocity. To asses whether the FW10 values are in fact tracing the higher velocity emission, as opposed to the strength of the chromospheric H$\alpha$ emission, we also computed the equivalent width (EW) of each stellar component and found them to be insensitive to the binary orbital phase. Typical EW values for the primary and secondary stars lay in the ranges 1.5--3 and 1--2.5 \AA, respectively\footnote{The sole exception is the profile at phase=0.663, which displays EWs of $\sim$ 3.5 and 5 \AA\ for the primary and secondary stars, respectively.}.

The significance of flux peaks and line broadening around the stellar eclipses, and the presence of short-duration flux dips preceding secondary eclipse (phase$\sim$0.4; see Fig. \ref{phot_spec_phase}, middle panel) are discussed in section \ref{sdfd} where we present a model of the inner regions of the binary system.

\section{Investigating the origin of the short-duration flux dips}
\label{sdfd}

In section~\ref{phot_var_sec} we sought to understand how much of the out-of-eclipse (OOE) variability could be attributed to starspots and found that the large scale structure in the system's 2008 and 2011/2012 optical light curves is consistent with the constructive and destructive interference of starspot signals at two slightly different periods. However, the residuals of our spot models show short-timescale variations that are not consistent with spot modulation and which show a tentative correlation with binary orbital phase, displaying peaks in IR flux around primary and secondary eclipse. These residual variations could arise from accretion-related processes and/or variable dust emission and obscuration. With respect to dust obscuration, the 2011/2012 CoRoT, 3.6 and 4.5 $\mu$m light curves all display simultaneous, short-duration flux dips preceding secondary eclipse (orange points, Fig.~\ref{spot_LCs} and phase$\sim$0.4 in Fig.~\ref{phot_spec_phase}, middle panel). These dips are not a permanent feature but are seen in four of the six orbital periods over which we have simultaneous photometry. There are also hints of much broader and shallower dips preceding primary eclipses, but these are harder to identify as they do not show such a disparate morphology to the large scale variations, although the dip at rHJD$\sim$5923 clearly shows a non-spot colour signature. In this section, we seek to identify the physical origin of the short-duration flux dips.

In an accreting binary, accretion streams are thought to flow from the circumbinary disk, through the corotating Lagrangian points, and onto the stars \citep[e.g.][]{deVal-Borro11}. 
A possible explanation for the flux dips could be that, as we are viewing the system close to edge on ($i \sim 85\degree$), these accretion streams partially occult one or both stars at certain phases of the binary orbit. In such a scenario, the accretion stream onto the primary could occult the binary before the primary star passes in front of the secondary, which would give dips preceding secondary eclipse, as observed. Similarly, the accretion stream onto the secondary could produce flux dips preceding primary eclipse. However, \citet{Shi12} find that a single dominant stream develops in 3D MHD simulations of an equal-mass binary and \citet{deVal-Borro11} find that mass is preferentially channelled towards the primary star in 2D hydrodynamic simulations for a non-equal mass system. In the scenario where the accretion stream onto the primary is dominant, this would cause stronger dips preceding secondary eclipse than for primary, as observed.
It is important to note that the amount of dust in accretion streams required to give short duration flux dips would not necessarily drive strong accretion, consistent with what we observe. Indeed, \citet{Bozhinova16} find evidence for variable extinction in young M dwarfs that do not possess IR excesses suggesting that very little dust is required to produce significant extinction signatures.

To ascertain whether accretion streams could cause the short-duration flux dips seen in CoRoT\,223992193, we set up a simple model of the binary system. We simulate the innermost regions of the circumbinary disk and the accretion flow in the central cavity with a custom-written particle code. To do this, we set up a ring of particles in Keplerian rotation around the binary centre of mass at an arbitrary distance beyond the circumbinary disk's theoretical inner edge (i.e. $>$0.1 AU).

We then give each particle a small radial velocity to force them to enter the cavity; this mimics turbulent processes in the disk that would cause particles to lose angular momentum and accrete onto the binary \citep{Shi12}. Each particle is subject to the gravitational potential from the stars, but we ignore pressure effects. This is a good approximation within the cavity because the density is very low, as has been confirmed by hydrodynamical simulations, which find that the motion of particles within the cavity is essentially ballistic. However, we note that hydrodynamical effects are important in the denser circumbinary disk and that our model will not capture these details. Nonetheless, our model is valid for analysing the accretion streams, which is our aim. Once a particle enters the cavity it can either be accreted onto one of the stars or accelerated onto a trajectory that sends it back into the circumbinary disk. In the latter case, we assume that the shock upon impact with the circumbinary disk circularises the particle's orbit and we therefore reset its velocity to the Keplerian value, with a small inward drift. 

\begin{figure}[t!]
  \centering  
  \includegraphics[width=\linewidth]{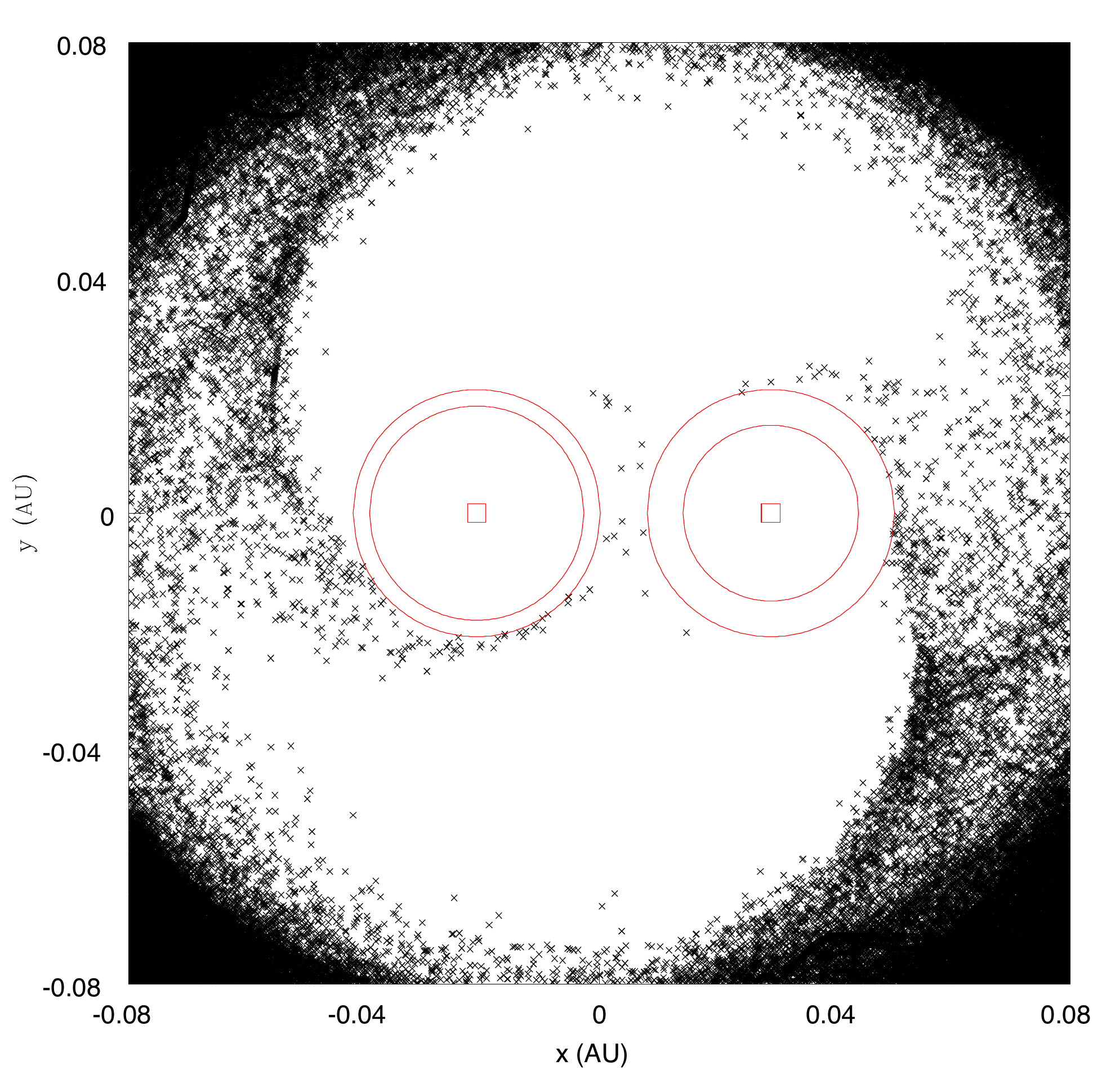}
   \caption{Snapshot of the accretion flow structure in the corotating frame of the binary, as output from our particle model. The red squares indicate the locations of the primary and secondary stars (left and right, respectively) and the black crosses show the locations of particles in the circumbinary disk and the accretion streams onto the two stars. The red circles delimit the inner and outer edges of where dust could reside in circumstellar disks: the inner edge represents the sublimation radius for each star and the outer edge the radius at which the disks are truncated by the other star. The binary clears out a slightly eccentric cavity of radius $\sim$0.08--0.1 AU ($\sim$17--22 $R_{\odot}$).}
   \label{geom}
\end{figure}

A snapshot of the resulting accretion flow is shown in Fig.~\ref{geom} in a frame co-rotating with the binary and centred on its centre of mass. The black crosses represent the positions of the particles and the red squares the position of the two stars, with the primary on the left. The red circles delimit rings in which, in principle, particles could be trapped in Keplerian motion around each of the stars, forming circumstellar disks. For each ring, the inner circle lies at the sublimation radius, within which no amorphous grains of dust can exist, and the outer circle corresponds to the radius at which circumstellar disks are tidally truncated by the other star \citep[roughly one third of the binary separation:][]{Paczynski77, Papaloizou77}. We have removed the particles that reach these rings, as we expect them to be continuously channeled onto the stars by the stellar magnetic fields, but it is plausible that some material exists in these rings. In our model, the binary clears out a slightly eccentric cavity of radius $\sim$0.08--0.1 AU ($\sim$17--22 $R_{\odot}$). We note that this eccentricity is consistent with 2D hydrodynamical and 3D MHD simulations \citep{Hanawa10,Shi12}.

The easiest way to interpret Fig.~\ref{geom} is to move ourselves around the plot. Primary eclipse occurs when we look directly from the right hand side of the plot (i.e. the secondary star on the right lies in front of the primary star on the left) and we move clockwise around the bottom until we are looking directly from the left hand side, which corresponds to secondary eclipse. As we do so, we see that as our line of sight moves around the bottom left quadrant our view to the primary star passes through the accretion stream onto the primary. Given the inclination and system scale, simple geometry predicts that the primary accretion stream would partially occult the primary star and its surrounding material (assuming negligible scale height as expected due to the low pressure). This obscuration could persist over the $\sim$$90\degree$ in phase prior to secondary eclipse, which corresponds to $\sim$1 day and is consistent with the typical duration of the obscuration events.

The model presented here has also been used to study emission from the inner edge of circumbinary discs and has been applied to CoRoT\,223992193 \citep{Terquem&Papaloizou16}. In addition, \citet{Terquem15} present 2D isothermal hydrodynamical simulations of CoRoT\,223992193. They find that the inclination and system geometry allow for the stars to be obscured by dust in the very inner regions of the system. For a given system scale, there is a fine balance between an inclination that is too high, such that the circumbinary disk obscures both stars completely, and too low, such that no obscuration occurs; CoRoT\,223992193 lies between these two options, giving stars that are visible but which can be partially (and time-variably) obscured due to the non-axisymmetric nature of the binary geometry. With material in the inner regions of the system, the general variations seen in the residuals of our 2011/2012 spot models (Figs.~\ref{spot_LCs}, \ref{col_mag} and \ref{phot_spec_phase}) can be understood as arising from variable obscuration and emission (the latter because this material would be close to the sublimation temperature of $T \sim 1500$ K and therefore contributing to the 3.6 and 4.5 $\mu$m bands). As our simple model does not capture the full complexity of the system it is difficult to be more quantitative. Indeed, in the 3D MHD simulations of \citet{Shi15}, structures in the circumbinary disk appear more complicated than in hydrodynamical simulations and it is not unreasonable to expect a similar complexity within the cavity for this system.

While our model is very simple, we have shown that the expected accretion flow geometry in the cavity of the circumbinary disk can reproduce short-duration flux dips before secondary eclipse not unlike those seen in the CoRoT and \emph{Spitzer} light curves. The accretion flow shown in Fig.~\ref{geom} is very similar to that obtained by \citet{Hanawa10}, who perform hydrodynamical simulations of a binary with a mass ratio similar to CoRoT\,223992193. We therefore note that simple particle models, such as we describe here, can offer a useful alternative to computationally expensive simulations when investigating low-density phenomena, such as accretion streams in binary systems.

Finally, we return to the observed increases in IR photometric flux and H$\alpha$ FW10 around primary and secondary eclipses. It may be possible to explain these in the context of the above model. When the two stars align along the line of sight (i.e. at eclipse), the accretion streams also broadly align. Peaks in IR photometric flux, therefore, could arise from the reprocessing of stellar irradiation by material in the accretions streams, which would appear brighter when a stream points towards us. The increases in H$\alpha$ FW10 can also be understood in a similar sense because when material in the accretion streams flows towards (away) from the observer any emission will be blue (red) shifted, thereby increasing the width of accretion-sensitive lines such as H$\alpha$.


\section{Conclusions and future work}
\label{conclusions}

CoRoT\,223992193 is the only known low-mass, pre-main sequence eclipsing binary to show evidence of a circumbinary disk. In seeking to understand what physical processes drive the observed photometric variability, we have shown that the large scale structure in the system's optical CoRoT light curves is broadly consistent with the constructive and destructive interference of starspot signals at two slightly different periods. Using the $v \sin i$ of both stars, we interpret this as the two stars having slightly different rotation periods: the primary is consistent with synchronisation and the secondary rotates slightly supersynchronously. Modelling the simultaneous 2011/2012 CoRoT and \emph{Spitzer} 3.6 and 4.5 $\mu$m light curves indicates that additional variability is present, most notably in the \emph{Spitzer} bands. 
There appears to be a tentative correlation between this additional variability and the binary orbital phase, with the system displaying increases in its IR flux around primary and secondary eclipse.
Given that the system's SED requires the presence of dust in the central cavity of the circumbinary disk, we investigated the effect of variable dust emission as well as dust obscuration and find that the remaining variability is consistent with a combination of these two phenomena.

Analysis of 15 medium resolution H$\alpha$ profiles spread throughout the binary orbit reveal an emission component associated with each star. The majority of this is consistent with chromospheric emission but some profiles also display higher velocity emission. This emission could be due to prominences, although we note that half of the secondary emission components display full widths at 10\% intensity > 270 km\,s$^{-1}$, which can also be interpreted as having an accretion-related origin. Similarly, simultaneous $u$ and $r$-band photometry obtained the month after the CoRoT and \emph{Spitzer} observations reveal a short-lived $u$-band excess consistent with either a stellar flare or an accretion hot spot. We need simultaneous, multiple emission line observations at high resolution to break the degeneracy between chromospheric and any accretion-related processes. 

The 2011/2012 CoRoT and \emph{Spitzer} light curves revealed short-duration flux dips that precede secondary eclipse (i.e. they occur broadly in phase with the binary orbit). 
We constructed a simple model of the binary and find that the accretion stream onto the primary star could partially obscure light from the central binary at the observed phases.

Determining the physical origins of photometric and spectroscopic variability in young binary systems is difficult due to the geometric complexity and hence many potential sources of variability. Even with high-precision photometry and medium-resolution spectroscopy we cannot unambiguously disentangle the different signals. To do so would require high-precision multi-band photometry with simultaneous high-resolution spectroscopic monitoring of multiple emission lines.

While low-mass, accreting binary systems have been the focus of much study, e.g. V4046 Sgr and DQ Tau, this is the first such system containing an eclipsing binary. The photometric and spectroscopic variations of CoRoT\,223992193 are consistent with the picture of two active stars possibly undergoing non-steady, low-level accretion; the system's very high inclination provides a new view of such variability. We hope to improve our understanding of this system in the future by obtaining multi-epoch monitoring of a range of emission lines. Looking to the future, the \emph{Kepler}/K2 mission is observing a number of nearby star forming regions and young open clusters, and could discover similar, brighter objects that can shed further light on the processes driving variability in young binary systems.


\begin{acknowledgements}

We thank the anonymous referee for their insightful and positive report.
We thank Cathie Clarke and Stefano Facchini for helpful discussions. This work was supported in part by the UK Science and Technology Facilities Council through studentship ST/J500641/1 (EG) and grant ST/G002266 (SA). JB thanks CNES for partial funding on this project. SHPA acknowledges support by CNPq, CAPES and Fapemig.
This work is based on observations obtained with numerous facilities: the \emph{CoRoT space mission} has been developed and operated by CNES, with contributions from Austria, Belgium, Brazil, ESA (RSSD and Science Program), Germany, and Spain; the \emph{Spitzer Space Telescope}, which is operated by the Jet Propulsion Laboratory, California Institute of Technology under a contract with NASA; \emph{ESO Telescopes} at the La Silla Paranal Observatory under programme ID 088.C-0239(A); \emph{MegaPrime/MegaCam, a joint project of CFHT and CEA/DAPNIA, at the Canada-France-Hawaii Telescope (CFHT)} which is operated by the National Research Council (NRC) of Canada, the Institut National des Sciences de l'Univers of the Centre National de la Recherche Scientifique (CNRS) of France, and the University of Hawaii; and the \emph{Nordic Optical Telescope}, operated by the Nordic Optical Telescope Scientific Association at the Observatorio del Roque de los Muchachos, La Palma, Spain, of the Instituto de Astrofisica de Canarias. The research leading to these results has received funding from the European Union Seventh Framework Programme (FP7/2013-2016) under grant agreement No. 312430 (OPTICON).

\end{acknowledgements}

\bibliographystyle{aa}
\bibliography{ref}

\begin{thebibliography}{81}
\expandafter\ifx\csname natexlab\endcsname\relax\def\natexlab#1{#1}\fi

\bibitem[{{Alencar} {et~al.}(2012){Alencar}, {Bouvier}, {Walter}, {Dougados},
  {Donati}, {Kurosawa}, {Romanova}, {Bonfils}, {Lima}, {Massaro}, {Ibrahimov},
  \& {Poretti}}]{Alencar12}
{Alencar}, S.~H.~P., {Bouvier}, J., {Walter}, F.~M., {et~al.} 2012, \aap, 541,
  A116

\bibitem[{{Alencar} {et~al.}(2003){Alencar}, {Melo}, {Dullemond}, {Andersen},
  {Batalha}, {Vaz}, \& {Mathieu}}]{Alencar03}
{Alencar}, S.~H.~P., {Melo}, C.~H.~F., {Dullemond}, C.~P., {et~al.} 2003, \aap,
  409, 1037

\bibitem[{{Artymowicz} \& {Lubow}(1994)}]{Artymowicz94}
{Artymowicz}, P. \& {Lubow}, S.~H. 1994, \apj, 421, 651

\bibitem[{{Artymowicz} \& {Lubow}(1996)}]{Artymowicz96}
{Artymowicz}, P. \& {Lubow}, S.~H. 1996, \apjl, 467, L77

\bibitem[{{Basri} {et~al.}(1997){Basri}, {Johns-Krull}, \& {Mathieu}}]{Basri97}
{Basri}, G., {Johns-Krull}, C.~M., \& {Mathieu}, R.~D. 1997, \aj, 114, 781

\bibitem[{{Billot} {et~al.}(2012){Billot}, {Morales-Calder{\'o}n}, {Stauffer},
  {Megeath}, \& {Whitney}}]{Billot12}
{Billot}, N., {Morales-Calder{\'o}n}, M., {Stauffer}, J.~R., {Megeath}, S.~T.,
  \& {Whitney}, B. 2012, \apjl, 753, L35

\bibitem[{{Bouvier}(2013)}]{Bouvier13b}
{Bouvier}, J. 2013, in EAS Publications Series, Vol.~62, EAS Publications
  Series, ed. P.~{Hennebelle} \& C.~{Charbonnel}, 143--168

\bibitem[{{Bouvier} {et~al.}(2007){Bouvier}, {Alencar}, {Boutelier},
  {Dougados}, {Balog}, {Grankin}, {Hodgkin}, {Ibrahimov}, {Kun}, {Magakian}, \&
  {Pinte}}]{Bouvier07}
{Bouvier}, J., {Alencar}, S.~H.~P., {Boutelier}, T., {et~al.} 2007, \aap, 463,
  1017

\bibitem[{{Bouvier} {et~al.}(1993){Bouvier}, {Cabrit}, {Fernandez}, {Martin},
  \& {Matthews}}]{Bouvier93}
{Bouvier}, J., {Cabrit}, S., {Fernandez}, M., {Martin}, E.~L., \& {Matthews},
  J.~M. 1993, \aap, 272, 176

\bibitem[{{Bouvier} {et~al.}(1999){Bouvier}, {Chelli}, {Allain}, {Carrasco},
  {Costero}, {Cruz-Gonzalez}, {Dougados}, {Fern{\'a}ndez}, {Mart{\'{\i}}n},
  {M{\'e}nard}, {Mennessier}, {Mujica}, {Recillas}, {Salas}, {Schmidt}, \&
  {Wichmann}}]{Bouvier99}
{Bouvier}, J., {Chelli}, A., {Allain}, S., {et~al.} 1999, \aap, 349, 619

\bibitem[{{Bozhinova} {et~al.}(2016){Bozhinova}, {Scholz}, \&
  {Eisl{\"o}ffel}}]{Bozhinova16}
{Bozhinova}, I., {Scholz}, A., \& {Eisl{\"o}ffel}, J. 2016, \mnras, 458, 3118

\bibitem[{{Claret} {et~al.}(2012){Claret}, {Hauschildt}, \& {Witte}}]{Claret12}
{Claret}, A., {Hauschildt}, P.~H., \& {Witte}, S. 2012, \aap, 546, A14

\bibitem[{{Cody} {et~al.}(2014){Cody}, {Stauffer}, {Baglin}, {Micela},
  {Rebull}, {Flaccomio}, {Morales-Calder{\'o}n}, {Aigrain}, {Bouvier},
  {Hillenbrand}, {Gutermuth}, {Song}, {Turner}, {Alencar}, {Zwintz},
  {Plavchan}, {Carpenter}, {Findeisen}, {Carey}, {Terebey}, {Hartmann},
  {Calvet}, {Teixeira}, {Vrba}, {Wolk}, {Covey}, {Poppenhaeger}, {G{\"u}nther},
  {Forbrich}, {Whitney}, {Affer}, {Herbst}, {Hora}, {Barrado}, {Holtzman},
  {Marchis}, {Wood}, {Medeiros Guimar{\~a}es}, {Lillo Box}, {Gillen},
  {McQuillan}, {Espaillat}, {Allen}, {D'Alessio}, \& {Favata}}]{Cody14}
{Cody}, A.~M., {Stauffer}, J., {Baglin}, A., {et~al.} 2014, \aj, 147, 82

\bibitem[{{de Val-Borro} {et~al.}(2011){de Val-Borro}, {Gahm}, {Stempels}, \&
  {Pepli{\'n}ski}}]{deVal-Borro11}
{de Val-Borro}, M., {Gahm}, G.~F., {Stempels}, H.~C., \& {Pepli{\'n}ski}, A.
  2011, \mnras, 413, 2679

\bibitem[{{Donati} {et~al.}(1999){Donati}, {Collier Cameron}, {Hussain}, \&
  {Semel}}]{Donati99}
{Donati}, J.-F., {Collier Cameron}, A., {Hussain}, G.~A.~J., \& {Semel}, M.
  1999, \mnras, 302, 437

\bibitem[{{Donati} {et~al.}(2008){Donati}, {Jardine}, {Gregory}, {Petit},
  {Paletou}, {Bouvier}, {Dougados}, {M{\'e}nard}, {Collier Cameron}, {Harries},
  {Hussain}, {Unruh}, {Morin}, {Marsden}, {Manset}, {Auri{\`e}re}, {Catala}, \&
  {Alecian}}]{Donati08}
{Donati}, J.-F., {Jardine}, M.~M., {Gregory}, S.~G., {et~al.} 2008, \mnras,
  386, 1234

\bibitem[{{Donati} {et~al.}(2000){Donati}, {Mengel}, {Carter}, {Marsden},
  {Collier Cameron}, \& {Wichmann}}]{Donati00}
{Donati}, J.-F., {Mengel}, M., {Carter}, B.~D., {et~al.} 2000, \mnras, 316, 699

\bibitem[{{Dorren}(1987)}]{Dorren87}
{Dorren}, J.~D. 1987, \apj, 320, 756

\bibitem[{{Duch{\^e}ne} \& {Kraus}(2013)}]{Duchene13}
{Duch{\^e}ne}, G. \& {Kraus}, A. 2013, \araa, 51, 269

\bibitem[{{Espaillat} {et~al.}(2011){Espaillat}, {Furlan}, {D'Alessio},
  {Sargent}, {Nagel}, {Calvet}, {Watson}, \& {Muzerolle}}]{Espaillat11}
{Espaillat}, C., {Furlan}, E., {D'Alessio}, P., {et~al.} 2011, \apj, 728, 49

\bibitem[{{Fern{\'a}ndez} {et~al.}(2004){Fern{\'a}ndez}, {Stelzer}, {Henden},
  {Grankin}, {Gameiro}, {Costa}, {Guenther}, {Amado}, \&
  {Rodriguez}}]{Fernandez04}
{Fern{\'a}ndez}, M., {Stelzer}, B., {Henden}, A., {et~al.} 2004, \aap, 427, 263

\bibitem[{{Flaherty} \& {Muzerolle}(2010)}]{Flaherty10}
{Flaherty}, K.~M. \& {Muzerolle}, J. 2010, \apj, 719, 1733

\bibitem[{{Fonseca} {et~al.}(2014){Fonseca}, {Alencar}, {Bouvier}, {Favata}, \&
  {Flaccomio}}]{Fonseca14}
{Fonseca}, N.~N.~J., {Alencar}, S.~H.~P., {Bouvier}, J., {Favata}, F., \&
  {Flaccomio}, E. 2014, \aap, 567, A39

\bibitem[{{Foreman-Mackey} {et~al.}(2013){Foreman-Mackey}, {Hogg}, {Lang}, \&
  {Goodman}}]{Foreman-Mackey13}
{Foreman-Mackey}, D., {Hogg}, D.~W., {Lang}, D., \& {Goodman}, J. 2013, \pasp,
  125, 306

\bibitem[{{Frandsen} \& {Lindberg}(1999)}]{Frandsen99}
{Frandsen}, S. \& {Lindberg}, B. 1999, in Astrophysics with the NOT, ed.
  H.~{Karttunen} \& V.~{Piirola}, 71

\bibitem[{{Gandolfi} {et~al.}(2013){Gandolfi}, {Parviainen}, {Fridlund},
  {Hatzes}, {Deeg}, {Frasca}, {Lanza}, {Prada Moroni}, {Tognelli}, {McQuillan},
  {Aigrain}, {Alonso}, {Antoci}, {Cabrera}, {Carone}, {Csizmadia}, {Djupvik},
  {Guenther}, {Jessen-Hansen}, {Ofir}, \& {Telting}}]{Gandolfi13}
{Gandolfi}, D., {Parviainen}, H., {Fridlund}, M., {et~al.} 2013, \aap, 557, A74

\bibitem[{{Gillen} {et~al.}(2014){Gillen}, {Aigrain}, {McQuillan}, {Bouvier},
  {Hodgkin}, {Alencar}, {Terquem}, {Southworth}, {Gibson}, {Cody}, {Lendl},
  {Morales-Calder{\'o}n}, {Favata}, {Stauffer}, \& {Micela}}]{Gillen14}
{Gillen}, E., {Aigrain}, S., {McQuillan}, A., {et~al.} 2014, \aap, 562, A50

\bibitem[{{Grankin}(1998)}]{Grankin98}
{Grankin}, K.~N. 1998, Astronomy Letters, 24, 497

\bibitem[{{Grankin} {et~al.}(2008){Grankin}, {Bouvier}, {Herbst}, \&
  {Melnikov}}]{Grankin08}
{Grankin}, K.~N., {Bouvier}, J., {Herbst}, W., \& {Melnikov}, S.~Y. 2008, \aap,
  479, 827

\bibitem[{{Grankin} {et~al.}(2007){Grankin}, {Melnikov}, {Bouvier}, {Herbst},
  \& {Shevchenko}}]{Grankin07}
{Grankin}, K.~N., {Melnikov}, S.~Y., {Bouvier}, J., {Herbst}, W., \&
  {Shevchenko}, V.~S. 2007, \aap, 461, 183

\bibitem[{{Gray}(1992)}]{Gray92}
{Gray}, D.~F. 1992, {The observation and analysis of stellar photospheres.},
  Vol.~20 (Camb. Astrophys. Ser.)

\bibitem[{{G{\"u}nther} \& {Kley}(2002)}]{Gunther02}
{G{\"u}nther}, R. \& {Kley}, W. 2002, \aap, 387, 550

\bibitem[{{Hanawa} {et~al.}(2010){Hanawa}, {Ochi}, \& {Ando}}]{Hanawa10}
{Hanawa}, T., {Ochi}, Y., \& {Ando}, K. 2010, \apj, 708, 485

\bibitem[{{Hatzes}(1995)}]{Hatzes95}
{Hatzes}, A.~P. 1995, \aj, 109, 350

\bibitem[{{Herbst} {et~al.}(1994){Herbst}, {Herbst}, {Grossman}, \&
  {Weinstein}}]{Herbst94}
{Herbst}, W., {Herbst}, D.~K., {Grossman}, E.~J., \& {Weinstein}, D. 1994, \aj,
  108, 1906

\bibitem[{{Husser} {et~al.}(2013){Husser}, {Wende-von Berg}, {Dreizler},
  {Homeier}, {Reiners}, {Barman}, \& {Hauschildt}}]{Husser13}
{Husser}, T.-O., {Wende-von Berg}, S., {Dreizler}, S., {et~al.} 2013, \aap,
  553, A6

\bibitem[{{Indebetouw} {et~al.}(2005){Indebetouw}, {Mathis}, {Babler}, {Meade},
  {Watson}, {Whitney}, {Wolff}, {Wolfire}, {Cohen}, {Bania}, {Benjamin},
  {Clemens}, {Dickey}, {Jackson}, {Kobulnicky}, {Marston}, {Mercer},
  {Stauffer}, {Stolovy}, \& {Churchwell}}]{Indebetouw05}
{Indebetouw}, R., {Mathis}, J.~S., {Babler}, B.~L., {et~al.} 2005, \apj, 619,
  931

\bibitem[{{Jackson} \& {Jeffries}(2013)}]{Jackson13}
{Jackson}, R.~J. \& {Jeffries}, R.~D. 2013, \mnras, 431, 1883

\bibitem[{{Jackson} \& {Jeffries}(2014)}]{Jackson14}
{Jackson}, R.~J. \& {Jeffries}, R.~D. 2014, \mnras, 441, 2111

\bibitem[{{Jackson} {et~al.}(2009){Jackson}, {Jeffries}, \&
  {Maxted}}]{Jackson09}
{Jackson}, R.~J., {Jeffries}, R.~D., \& {Maxted}, P.~F.~L. 2009, \mnras, 399,
  L89

\bibitem[{{Jensen} {et~al.}(2007){Jensen}, {Dhital}, {Stassun}, {Patience},
  {Herbst}, {Walter}, {Simon}, \& {Basri}}]{Jensen07}
{Jensen}, E.~L.~N., {Dhital}, S., {Stassun}, K.~G., {et~al.} 2007, \aj, 134,
  241

\bibitem[{{Jones} {et~al.}(1996){Jones}, {Fischer}, \& {Stauffer}}]{Jones96}
{Jones}, B.~F., {Fischer}, D.~A., \& {Stauffer}, J.~R. 1996, \aj, 112, 1562

\bibitem[{{Joy}(1945)}]{Joy45}
{Joy}, A.~H. 1945, \apj, 102, 168

\bibitem[{{King} {et~al.}(2000){King}, {Soderblom}, {Fischer}, \&
  {Jones}}]{King00}
{King}, J.~R., {Soderblom}, D.~R., {Fischer}, D., \& {Jones}, B.~F. 2000, \apj,
  533, 944

\bibitem[{{Long} {et~al.}(2007){Long}, {Romanova}, \& {Lovelace}}]{Long07}
{Long}, M., {Romanova}, M.~M., \& {Lovelace}, R.~V.~E. 2007, \mnras, 374, 436

\bibitem[{{Long} {et~al.}(2008){Long}, {Romanova}, \& {Lovelace}}]{Long08}
{Long}, M., {Romanova}, M.~M., \& {Lovelace}, R.~V.~E. 2008, \mnras, 386, 1274

\bibitem[{{Makidon} {et~al.}(2004){Makidon}, {Rebull}, {Strom}, {Adams}, \&
  {Patten}}]{Makidon04}
{Makidon}, R.~B., {Rebull}, L.~M., {Strom}, S.~E., {Adams}, M.~T., \& {Patten},
  B.~M. 2004, \aj, 127, 2228

\bibitem[{{Mathieu} {et~al.}(1997){Mathieu}, {Stassun}, {Basri}, {Jensen},
  {Johns-Krull}, {Valenti}, \& {Hartmann}}]{Mathieu97}
{Mathieu}, R.~D., {Stassun}, K., {Basri}, G., {et~al.} 1997, \aj, 113, 1841

\bibitem[{{McGinnis} {et~al.}(2015){McGinnis}, {Alencar}, {Guimar{\~a}es},
  {Sousa}, {Stauffer}, {Bouvier}, {Rebull}, {Fonseca}, {Venuti}, {Hillenbrand},
  {Cody}, {Teixeira}, {Aigrain}, {Favata}, {F{\H u}r{\'e}sz}, {Vrba},
  {Flaccomio}, {Turner}, {Gameiro}, {Dougados}, {Herbst},
  {Morales-Calder{\'o}n}, \& {Micela}}]{McGinnis15}
{McGinnis}, P.~T., {Alencar}, S.~H.~P., {Guimar{\~a}es}, M.~M., {et~al.} 2015,
  \aap, 577, A11

\bibitem[{{McKay} {et~al.}(1979){McKay}, {Beckman}, \& {Conover}}]{McKay79}
{McKay}, M.~D., {Beckman}, R.~J., \& {Conover}, W.~J. 1979, Technometrics, 21,
  239

\bibitem[{{Morales-Calder{\'o}n} {et~al.}(2011){Morales-Calder{\'o}n},
  {Stauffer}, {Hillenbrand}, {Gutermuth}, {Song}, {Rebull}, {Plavchan},
  {Carpenter}, {Whitney}, {Covey}, {Alves de Oliveira}, {Winston},
  {McCaughrean}, {Bouvier}, {Guieu}, {Vrba}, {Holtzman}, {Marchis}, {Hora},
  {Wasserman}, {Terebey}, {Megeath}, {Guinan}, {Forbrich}, {Hu{\'e}lamo},
  {Riviere-Marichalar}, {Barrado}, {Stapelfeldt}, {Hern{\'a}ndez}, {Allen},
  {Ardila}, {Bayo}, {Favata}, {James}, {Werner}, \&
  {Wood}}]{Morales-Calderon11}
{Morales-Calder{\'o}n}, M., {Stauffer}, J.~R., {Hillenbrand}, L.~A., {et~al.}
  2011, \apj, 733, 50

\bibitem[{{O'Neal} {et~al.}(1998){O'Neal}, {Neff}, \& {Saar}}]{ONeal98}
{O'Neal}, D., {Neff}, J.~E., \& {Saar}, S.~H. 1998, \apj, 507, 919

\bibitem[{{O'Neal} {et~al.}(2004){O'Neal}, {Neff}, {Saar}, \&
  {Cuntz}}]{ONeal04}
{O'Neal}, D., {Neff}, J.~E., {Saar}, S.~H., \& {Cuntz}, M. 2004, \aj, 128, 1802

\bibitem[{{Paczynski}(1977)}]{Paczynski77}
{Paczynski}, B. 1977, \apj, 216, 822

\bibitem[{{Papaloizou} \& {Pringle}(1977)}]{Papaloizou77}
{Papaloizou}, J. \& {Pringle}, J.~E. 1977, \mnras, 181, 441

\bibitem[{{Parks} {et~al.}(2014){Parks}, {Plavchan}, {White}, \&
  {Gee}}]{Parks14}
{Parks}, J.~R., {Plavchan}, P., {White}, R.~J., \& {Gee}, A.~H. 2014, \apjs,
  211, 3

\bibitem[{{Petrov} {et~al.}(1994){Petrov}, {Shcherbakov}, {Berdyugina},
  {Shevchenko}, {Grankin}, \& {Melnikov}}]{Petrov94}
{Petrov}, P.~P., {Shcherbakov}, V.~A., {Berdyugina}, S.~V., {et~al.} 1994,
  \aaps, 107, 9

\bibitem[{{Rasmussen} \& {Williams}(2006)}]{Rasmussen06}
{Rasmussen}, C.~E. \& {Williams}, C.~K.~I. 2006, Gaussian Processes for Machine
  Learning, MIT Press

\bibitem[{{Rebull} {et~al.}(2014){Rebull}, {Cody}, {Covey}, {G{\"u}nther},
  {Hillenbrand}, {Plavchan}, {Poppenhaeger}, {Stauffer}, {Wolk}, {Gutermuth},
  {Morales-Calder{\'o}n}, {Song}, {Barrado}, {Bayo}, {James}, {Hora}, {Vrba},
  {Alves de Oliveira}, {Bouvier}, {Carey}, {Carpenter}, {Favata}, {Flaherty},
  {Forbrich}, {Hernandez}, {McCaughrean}, {Megeath}, {Micela}, {Smith},
  {Terebey}, {Turner}, {Allen}, {Ardila}, {Bouy}, \& {Guieu}}]{Rebull14}
{Rebull}, L.~M., {Cody}, A.~M., {Covey}, K.~R., {et~al.} 2014, \aj, 148, 92

\bibitem[{{Roettenbacher} {et~al.}(2013){Roettenbacher}, {Monnier}, {Harmon},
  {Barclay}, \& {Still}}]{Roettenbacher13}
{Roettenbacher}, R.~M., {Monnier}, J.~D., {Harmon}, R.~O., {Barclay}, T., \&
  {Still}, M. 2013, \apj, 767, 60

\bibitem[{{Romanova} {et~al.}(2008){Romanova}, {Kulkarni}, \&
  {Lovelace}}]{Romanova08}
{Romanova}, M.~M., {Kulkarni}, A.~K., \& {Lovelace}, R.~V.~E. 2008, \apjl, 673,
  L171

\bibitem[{{Romanova} {et~al.}(2011){Romanova}, {Ustyugova}, {Koldoba}, \&
  {Lovelace}}]{Romanova11}
{Romanova}, M.~M., {Ustyugova}, G.~V., {Koldoba}, A.~V., \& {Lovelace},
  R.~V.~E. 2011, \mnras, 416, 416

\bibitem[{{Romanova} {et~al.}(2013){Romanova}, {Ustyugova}, {Koldoba}, \&
  {Lovelace}}]{Romanova13}
{Romanova}, M.~M., {Ustyugova}, G.~V., {Koldoba}, A.~V., \& {Lovelace},
  R.~V.~E. 2013, \mnras, 430, 699

\bibitem[{{Romanova} {et~al.}(2003){Romanova}, {Ustyugova}, {Koldoba}, {Wick},
  \& {Lovelace}}]{Romanova03}
{Romanova}, M.~M., {Ustyugova}, G.~V., {Koldoba}, A.~V., {Wick}, J.~V., \&
  {Lovelace}, R.~V.~E. 2003, \apj, 595, 1009

\bibitem[{{Schlegel} {et~al.}(1998){Schlegel}, {Finkbeiner}, \&
  {Davis}}]{Schlegel98}
{Schlegel}, D.~J., {Finkbeiner}, D.~P., \& {Davis}, M. 1998, \apj, 500, 525

\bibitem[{{Shang} {et~al.}(2002){Shang}, {Glassgold}, {Shu}, \&
  {Lizano}}]{Shang02}
{Shang}, H., {Glassgold}, A.~E., {Shu}, F.~H., \& {Lizano}, S. 2002, \apj, 564,
  853

\bibitem[{{Shi} \& {Krolik}(2015)}]{Shi15}
{Shi}, J.-M. \& {Krolik}, J.~H. 2015, \apj, 807, 131

\bibitem[{{Shi} {et~al.}(2012){Shi}, {Krolik}, {Lubow}, \& {Hawley}}]{Shi12}
{Shi}, J.-M., {Krolik}, J.~H., {Lubow}, S.~H., \& {Hawley}, J.~F. 2012, \apj,
  749, 118

\bibitem[{{Skelly} {et~al.}(2009){Skelly}, {Unruh}, {Barnes}, {Lawson},
  {Donati}, \& {Collier Cameron}}]{Skelly09}
{Skelly}, M.~B., {Unruh}, Y.~C., {Barnes}, J.~R., {et~al.} 2009, \mnras, 399,
  1829

\bibitem[{{Stassun} {et~al.}(2014){Stassun}, {Feiden}, \& {Torres}}]{Stassun14}
{Stassun}, K.~G., {Feiden}, G.~A., \& {Torres}, G. 2014, \nar, 60, 1

\bibitem[{{Stassun} {et~al.}(2012){Stassun}, {Kratter}, {Scholz}, \&
  {Dupuy}}]{Stassun12}
{Stassun}, K.~G., {Kratter}, K.~M., {Scholz}, A., \& {Dupuy}, T.~J. 2012, \apj,
  756, 47

\bibitem[{{Stauffer} {et~al.}(2014){Stauffer}, {Cody}, {Baglin}, {Alencar},
  {Rebull}, {Hillenbrand}, {Venuti}, {Turner}, {Carpenter}, {Plavchan},
  {Findeisen}, {Carey}, {Terebey}, {Morales-Calder{\'o}n}, {Bouvier}, {Micela},
  {Flaccomio}, {Song}, {Gutermuth}, {Hartmann}, {Calvet}, {Whitney}, {Barrado},
  {Vrba}, {Covey}, {Herbst}, {Furesz}, {Aigrain}, \& {Favata}}]{Stauffer14}
{Stauffer}, J., {Cody}, A.~M., {Baglin}, A., {et~al.} 2014, \aj, 147, 83

\bibitem[{{Stauffer} {et~al.}(2015){Stauffer}, {Cody}, {McGinnis}, {Rebull},
  {Hillenbrand}, {Turner}, {Carpenter}, {Plavchan}, {Carey}, {Terebey},
  {Morales-Calder{\'o}n}, {Alencar}, {Bouvier}, {Venuti}, {Hartmann}, {Calvet},
  {Micela}, {Flaccomio}, {Song}, {Gutermuth}, {Barrado}, {Vrba}, {Covey},
  {Padgett}, {Herbst}, {Gillen}, {Lyra}, {Medeiros Guimaraes}, {Bouy}, \&
  {Favata}}]{Stauffer15}
{Stauffer}, J., {Cody}, A.~M., {McGinnis}, P., {et~al.} 2015, \aj, 149, 130

\bibitem[{{Sung} {et~al.}(2009){Sung}, {Stauffer}, \& {Bessell}}]{Sung09}
{Sung}, H., {Stauffer}, J.~R., \& {Bessell}, M.~S. 2009, \aj, 138, 1116

\bibitem[{{Symington} {et~al.}(2005){Symington}, {Harries}, {Kurosawa}, \&
  {Naylor}}]{Symington05}
{Symington}, N.~H., {Harries}, T.~J., {Kurosawa}, R., \& {Naylor}, T. 2005,
  \mnras, 358, 977

\bibitem[{{Telting} {et~al.}(2014){Telting}, {Avila}, {Buchhave}, {Frandsen},
  {Gandolfi}, {Lindberg}, {Stempels}, {Prins}, \& {NOT staff}}]{Telting14}
{Telting}, J.~H., {Avila}, G., {Buchhave}, L., {et~al.} 2014, Astronomische
  Nachrichten, 335, 41

\bibitem[{{Terquem} \& {Papaloizou}(2016)}]{Terquem&Papaloizou16}
{Terquem}, C. \& {Papaloizou}, J.~C.~B. 2016, \mnras

\bibitem[{{Terquem} {et~al.}(2015){Terquem}, {S{\o}rensen-Clark}, \&
  {Bouvier}}]{Terquem15}
{Terquem}, C., {S{\o}rensen-Clark}, P.~M., \& {Bouvier}, J. 2015, \mnras, 454,
  3472

\bibitem[{{Venuti} {et~al.}(2014){Venuti}, {Bouvier}, {Flaccomio}, {Alencar},
  {Irwin}, {Stauffer}, {Cody}, {Teixeira}, {Sousa}, {Micela}, {Cuillandre}, \&
  {Peres}}]{Venuti14}
{Venuti}, L., {Bouvier}, J., {Flaccomio}, E., {et~al.} 2014, \aap, 570, A82

\bibitem[{{Venuti} {et~al.}(2015){Venuti}, {Bouvier}, {Irwin}, {Stauffer},
  {Hillenbrand}, {Rebull}, {Cody}, {Alencar}, {Micela}, {Flaccomio}, \&
  {Peres}}]{Venuti15}
{Venuti}, L., {Bouvier}, J., {Irwin}, J., {et~al.} 2015, \aap, 581, A66

\bibitem[{{White} \& {Basri}(2003)}]{White03}
{White}, R.~J. \& {Basri}, G. 2003, \apj, 582, 1109

\end{thebibliography}

\begin{landscape}

\begin{table}[p]
  \begin{center}
  \caption[]{ Parameters of the three-Gaussian model used to simultaneously fit the H$\alpha$ profiles. Each profile was modelled as the sum of three gaussians, which represent emission associated with the nebula, the primary star and the secondary star, respectively. The widths of each Gaussian, along with the radial velocity of the nebula component, were jointly fit for using all profiles, and are shown at the bottom of the table. }
  \label{Halpha_par}
  \begin{tabular}{ l l @{\hskip 8mm}l @{\hskip 8mm}l l l @{\hskip 10mm}l l l }
    \hline
    \hline
    \noalign{\smallskip}
    Phase   &   HJD   &   Nebula   &   \multicolumn{3}{c}{Primary ~~~~~~~~~}   &   \multicolumn{3}{c}{Secondary ~~~~}   \\
      &  &  \emph{Amplitude}  &  \emph{Amplitude}  &  \emph{~~~RV}  &  \emph{$\Delta$ RV *}  &  \emph{Amplitude}  &  \emph{~~~RV}  &  \emph{$\Delta$ RV *} \\
      &  &  &  &  (km\,s$^{-1}$)&  (km\,s$^{-1}$) &  &  (km\,s$^{-1}$)  &  (km\,s$^{-1}$)  \\
    \noalign{\smallskip}
    \hline
    \noalign{\smallskip}
    
0.077  &  2455939.65778  &  $ 2.821\pm0.017$  &  $ 0.839\pm0.017$  &  $-28.68\pm0.97$         &  $\,\,\,-0.8$         &  $ 0.332\pm0.010$  &  $\,\,\,\,60.9\pm2.3$      &  $\,\,\,\,23.3$   \\ [0.5ex]
0.117  &  2455943.68750  &  $ 2.382\pm0.018$  &  $ 0.989\pm0.010$  &  $-39.36\pm0.63$         &  $ \,\,\,\,\,\,\,1.0$  &  $ 0.467\pm0.007$  &  $\,\,\,\,68.8\pm1.2$      &  $\,\,\,\,14.4$   \\ [0.5ex]
0.331  &  2455940.64342  &  $ 0.905\pm0.011$  &  $ 1.195\pm0.006$  &  $-53.04\pm0.32$         &  $\,\,\,-0.1$          &  $ 0.450\pm0.004$  &  $\,\,\,\,66.94\pm0.84$  &  $\,\,\,-4.4$   \\ [0.5ex]
0.341  &  2455940.67976  &  $ 0.805\pm0.010$  &  $ 1.132\pm0.005$  &  $-50.51\pm0.33$         &  $\,\,\,\,\,\,\,0.6$   &  $ 0.418\pm0.004$  &  $\,\,\,\,66.49\pm0.85$  &  $\,\,\,-2.4$   \\ [0.5ex]
0.362  &  2455944.63515  &  $ 1.139\pm0.011$  &  $ 0.966\pm0.007$  &  $-49.29\pm0.45$         &  $\,\,\,-2.9$          &  $ 0.629\pm0.005$  &  $\,\,\,\,60.05\pm0.78$  &  $\,\,\,-2.6$   \\ [0.5ex]
0.380  &  2455979.57946  &  $ 1.103\pm0.011$  &  $ 0.863\pm0.007$  &  $-39.53\pm0.47$         &  $\,\,\,\,\,\,\,2.0$   &  $ 0.339\pm0.005$  &  $\,\,\,\,61.0\pm1.2$      &  $\,\,\,\,\,\,\,5.0$   \\ [0.5ex]
0.411  &  2455917.70423  &  $ 2.960\pm0.016$  &  $ 0.743\pm0.014$  &  $-46.4\pm1.0$             &  $-14.1$              &  $ 0.583\pm0.010$  &  $\,\,\,\,45.9\pm1.5$      &  $\,\,\,\,\,\,\,2.3$   \\ [0.5ex]
0.594  &  2455941.66163  &  $ 1.242\pm0.016$  &  $ 0.797\pm0.010$  &  $\,\,\,\,41.52\pm0.85$  &  $\,\,\,\,\,\,\,7.7$   &  $ 0.487\pm0.006$  &  $-72.0\pm1.2$             &  $-26.3$   \\ [0.5ex]
0.628  &  2455945.66720  &  $ 0.836\pm0.010$  &  $ 1.061\pm0.005$  &  $\,\,\,\,47.41\pm0.36$  &  $\,\,\,\,\,\,\,3.7$   &  $ 0.673\pm0.004$  &  $-76.17\pm0.59$         &  $-17.2$   \\ [0.5ex]
0.663  &  2455918.68222  &  $ 2.393\pm0.013$  &  $ 1.777\pm0.007$  &  $\,\,\,\,57.56\pm0.28$  &  $\,\,\,\,\,\,\,5.7$   &  $ 1.003\pm0.005$  &  $-69.73\pm0.63$         &  $\,\,\,\,\,\,\,0.3$   \\ [0.5ex]
0.700  &  2455922.69870  &  $ 2.531\pm0.011$  &  $ 0.938\pm0.005$  &  $\,\,\,\,62.18\pm0.40$  &  $\,\,\,\,\,\,\,4.5$   &  $ 0.726\pm0.005$  &  $-91.66\pm0.57$         &  $-13.9$   \\ [0.5ex]
0.880  &  2455977.63971  &  $ 1.114\pm0.012$  &  $ 0.763\pm0.008$  &  $\,\,\,\,44.00\pm0.57$  &  $\,\,\,\,\,\,\,2.7$   &  $ 0.523\pm0.006$  &  $-55.6\pm1.1$             &  $\,\,\,\,\,\,\,0.1$   \\ [0.5ex]
0.882  &  2455915.65584  &  $ 2.164\pm0.014$  &  $ 0.741\pm0.007$  &  $\,\,\,\,50.66\pm0.67$  &  $\,\,\,\,10.0$       &  $ 0.454\pm0.006$  &  $-72.2\pm1.1$             &  $-17.4$   \\ [0.5ex]
0.890  &  2455946.68154  &  $ 2.808\pm0.015$  &  $ 0.927\pm0.010$  &  $\,\,\,\,49.24\pm0.64$  &  $\,\,\,\,10.8$       &  $ 0.716\pm0.007$  &  $-56.7\pm1.1$             &  $\,\,\,-4.9$   \\ [0.5ex]
0.906  &  2455981.61509  &  $ 1.584\pm0.015$  &  $ 0.766\pm0.014$  &  $\,\,\,\,37.72\pm0.88$  &  $\,\,\,\,\,\,\,4.1$   &  $ 0.416\pm0.011$  &  $-47.0\pm2.0$             &  $\,\,\,-1.7$   \\ [0.5ex]

\noalign{\smallskip}
\noalign{\smallskip}

 \multicolumn{9}{c}{\emph{Parameters jointly fit from all profiles}} \\
 \noalign{\smallskip}
    &      &   Nebula   &   \multicolumn{3}{c}{Primary ~~~~~~~~~}   &   \multicolumn{3}{c}{Secondary ~~~~} \\
 \noalign{\smallskip}
\multicolumn{2}{c}{\emph{Width (km\,s$^{-1}$)}}  &  $13.978\pm0.036$  &  \multicolumn{3}{c}{$43.50\pm0.16$}  &  \multicolumn{3}{c}{$53.83\pm0.47$}  \\ 
\noalign{\smallskip}
\multicolumn{2}{c}{\emph{RV  (km\,s$^{-1}$)}}  &  $-5.371\pm0.026$  &  &  &  &    \\

    \noalign{\smallskip}
    \hline
  \end{tabular}
  \end{center}
  \begin{list}{}{}
  \item[*] $\Delta$\,RV = RV$_{\rm{H\alpha}}$ -- RV$_{\rm{orbit}}$ (see section 3.2.1 of Paper 1 for details on the orbit determination).
  \end{list}
\end{table}

\end{landscape}

\end{document}